\documentclass[10pt]{iopart}
\usepackage{iopams}  
\usepackage{epsfig}
\begin{document}
\vspace*{-2.7cm}
\title[]
{CASTOR: Centauro And Strange Object Research in nucleus-nucleus
collisions at LHC\\\vspace*{5mm}}

\begin{center}
\begin{large}
{\em Ewa G\l{}adysz-Dziadu\'s for the CASTOR group}
\end{large}
\end{center}
\author{
A~L~S~Angelis$^1$,
X~Aslanoglou$^2$,
J~Bartke$^3$,
K~Chileev$^4$,
E~G\l{}adysz-Dziadu\'s$^3$~\footnote[7]{Corresponding author 
(ewa.gladysz@ifj.edu.pl)\\
Further information and complete bibliography at
http://home.cern.ch/angelis/castor/Welcome.html
},
M~Golubeva$^4$,
F~Guber$^4$,
T~Karavitcheva$^4$,
Y~V~Kharlov$^5$,
A~B~Kurepin$^4$,
G~Mavromanolakis$^1$,
A~D~Panagiotou$^1$,
S~A~Sadovsky$^5$
V~V~Tiflov$^4$,
and
Z~W\l{}odarczyk$^6$
}

\address{$^1$ Nuclear and Particle Physics Division, University of Athens,
              Athens, Greece.}
\address{$^2$ Department of Physics, the University of Ioannina,
              Ioannina, Greece.}
\address{$^3$ Institute of Nuclear Physics, Cracow, Poland.}
\address{$^4$ Institute for Nuclear Research, Moscow, Russia.}
\address{$^5$ Institute for High Energy Physics, Protvino, Russia.}
\address{$^6$ Institute of Physics, Pedagogical University, Kielce, Poland.}

\vspace*{-0.3cm}

\begin{abstract}

We describe the CASTOR detector designed to probe the very
forward, baryon-rich rapidity region in nucleus-nucleus collisions at the
LHC.
We present a phenomenological model describing the formation of a 
QGP fireball in high baryochemical potential
environment, and its subsequent
decay into baryons and possibly strangelets.
The model explains the Centauro events observed in cosmic rays and the
long-penetrating component frequently accompanying them, and makes
predictions for the LHC.
 Simulations of Centauro-type events by means of
our Monte-Carlo event generator CNGEN were done. To study the response
 of the apparatus to new effects,
  different exotic species (DCC clusters, Centauros, strangelets 
  and so--called mixed events produced by baryons and
strangelets being the remnants of the Centauro fireball explosion) were
 passed through the deep
calorimeter. The energy deposition pattern in the calorimeter appears to
be a new clear signature of the QGP state.
\end{abstract}
\vspace*{-5mm}
%Uncomment for PACS numbers title message
%\vspace*{-20pt}
%\pacs{12.38.Mh, 25.75.-q, 96.40.-z}

% Comment out if separate title page not required
% \maketitle

\enlargethispage{60pt}
\section{Introduction}
\label{sec:intro}

The motivation to study the very forward phase space in nucleus-nucleus 
collisions
at the LHC stems from the potentially very rich field of new phenomena
which can  be produced in an environment of very high baryochemical
potential.
The study of this baryon-dense region in the laboratory will provide
important information for the understanding of a Quark--Gluon Plasma (QGP)
state at relatively
low temperatures, with different properties from the one in the higher
temperature baryon-free region around mid-rapidity, which could exist in
the core of neutron stars.
Although there are serious technical difficulties in doing calculations
for a high baryochemical potential environment, many physicists agree
that a lot of interesting phenomena predicted by theory and/or
announced by experiments should  appear in this
region.

 In particular, some theoretical considerations suggest
\cite{ref:end_point} that
 the phase diagram features a critical endpoint E ($T_{E} \simeq 160$ MeV
and 
 $\mu_{E} \simeq$ 725 MeV) 
at which the line
of the first order phase transition ($\mu > \mu_{E}$ and $T < T_{E}$)
ends. At this point the phase transition becomes of second order and long
wavelength fluctuations appear. Passing close enough to
this critical endpoint 
should have characteristic experimental consequences.
Since one can miss the critical
point on either of two sides a nonmonotonic
dependence of the control parameters should be expected.

Other theoretical  ideas
  attracting a lot of attention are:
  colour superconductive state at finite baryon density
\cite{ref:superconductivity} or skyrmions \cite{ref:skyrmions} - coherent
states of baryons possibly produced by a DCC. 
 Also strangelets,  droplets of  strange
quark matter are predicted to be formed  from the Quark Gluon Plasma,
predominantly in a high baryochemical potential environment
\cite{ref:gre}.
Heavy flavour \cite{ref:Tien-Shan} and Super Heavy Particles
\cite{ref:heavy_part} production is
expected to dominate in
the  forward rapidity region.
 
 It is especially important to note that  high energy cosmic
ray
interactions
show the existence of  the wide spectrum of exotic events (Centauros,
 Mini--Centauros, Chirons, Geminions, Halo-type events etc.
\cite{ref:lat,ref:ewa})   
 observed
at forward rapidities.
These so--called Centauro species  reveal many  surprising
features, such as:
 abnormal hadron dominance,
  transverse momentum of produced particles  much higher than
that
      observed in ``normal'' interactions,
  the existence of mini--clusters etc..  Besides that they are
very frequently
 connected with the so-called {\em long-flying (penetrating) component}
 \cite{ref:ewa}-\cite{ref:Baradzei}.

These  anomalies  are observed at energies above $\sim
10^{15}$ eV and  are not rare occurence but they manifest
themselves at about 
5\% level.
 It is widely believed that Centauro related phenomena could not be due
to
any kind of statistical fluctuation in the hadronic content of normal
events and they have until now defied all attempts at explanation in terms
of
conventional physics~\cite{ref:tam}.
 Instead  many  unconventional models
 have been proposed.
 Some of them (e.g. \cite{ref:Bjorken_Cen}) assume that 
exotic objects
of unknown
origin are
present in the primary cosmic ray spectrum and they are seen as
Centauros during their penetration
through the  atmosphere. Others assume that  exotic fireballs
are
produced in
extremely high energy
hadron-hadron (e.g. \cite{ref:Goulianos}) or nucleus-nucleus
(e.g. \cite{ref:pa1,ref:asp}) interactions.
Other unconventional  attempts, as for example a DCC scenario
\cite{ref:DCC_Chac_1}
 or
 the color-sextet quark model
\cite{ref:White}, based on Pomeron physics in QCD were also developed.
 The widespread opinion
that the  likely mechanism for Centauro production is the
formation of a
quark-gluon  plasma  was incorporated in a lot of
proposed models. But only  the model of the strange
quark--matter
 fireball \cite{ref:pa1,ref:asp,ref:gl1} explains simultaneously 
 both the
main
features of the Centauro-like events and the strongly
penetrating component accompanying them.
Both the experimental  characteristics of Centauro--related species and
the model predictions
indicate the forward rapidity region as the most favourable  place for
 production and detection of such anomalous phenomena.

The LHC will be the first accelerator to effectively probe the very high 
energy cosmic ray domain, where cosmic ray experiments have detected
numerous very unusual events. These events
 may be produced and studied at the LHC in
controlled conditions.
 Majority of the present and future nucleus-nucleus experiments concern
 the exploration of the baryon-free region and  
  ''midrapidity physics''.  Already several years ago we announced the
necessity to investigate the forward
rapidity region in future heavy ion experiments 
 \cite{ref:midrap_physics}. 
 A small collaboration  has been formed and
 the CASTOR detector, a unique experimental design
 to probe the very forward rapidity region  in
 nucleus-nucleus collisions
at the LHC and  to complement the CERN heavy ion physics program  
pursued essentially in the baryon-free midrapidity region,
 has been proposed \cite{ref:Castor}.
In order to illustrate the detector's sensitivity to new effects
we have done  simulations of Centauro-type events by means of
our Monte-Carlo event generator CNGEN \cite{ref:cng}.
 The simulated
 Centauro  events have characteristics manifestly different
from those
 predicted by ``classical'' (e.g. HIJING) generators
\cite{ref:ewa,ref:mix}.
 The different exotic species  were also followed through the deep
calorimeter by means of modified GEANT 3.21
\cite{ref:ewa,ref:mix,ref:aa2}. We
simulated transition
curves
produced  in the CASTOR calorimeter by: DCC clusters
(both neutral and charged), Centauros, strangelets (both
 stable and unstable) and so--called mixed events produced by baryons and
strangelets as the remnants of the Centauro fireball explosion.
 To study the sensitivity of the
calorimeter to abnormally penetrating objects a neural network technique 
was also developed~\cite{ref:SQM2001}.
 Different exotic phenomena
 give different
 energy deposition patterns and can be well  distinguished
from the
 usual events as well as  from one another.

%\vspace*{20pt}

\section{Centauro--like phenomena in cosmic ray experiments}
\label{sec:centexp}

  Centauro related phenomena were discovered and have been
  analysed  
 in emulsion chamber experiments investigating  cosmic ray interactions
 at  the high  mountain laboratories  at Mt. Chacaltaya (5200
m above see level
% (540 g/cm$^{2}$)
) and Pamirs ($\sim$ 4300
%(596 g/cm$^{2}$)
or 4900 m above sea level). Both the experimental aspects and the model 
explanations have been presented in the recent review \cite{ref:ewa}.

\begin{figure}[h]
\vspace*{-1cm}
%\begin{minipage}{9cm}
\begin{minipage}{7cm}
\mbox{
\epsfxsize=200pt
\epsfbox[25 90 600 850]{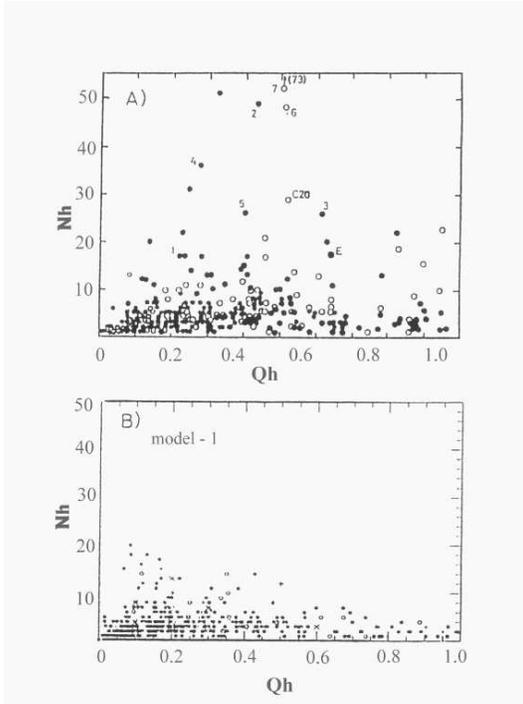}}
\end{minipage}
\begin{minipage}{8cm}
\vspace{3cm}
\caption[{\scriptsize (A) $N_{h}-Q_{h}$ diagram of families
  detected in Pamir,
   Chacaltaya and Pamir-Joint chambers.}]
{{\scriptsize (A) $N_{h}-Q_{h}$ diagram of families
  detected in Pamir,
   Chacaltaya and Pamir-Joint chambers,
  (B) The same for the simulated families. Different marks
 signify the different primary cosmic-ray nuclei: ($\bullet$) proton,
 ($\circ$)  $\alpha$, ($\diamond$)  CNO, ($\times$)  heavy, (+)  Fe,
  \cite{ref:Baradzei}.}}
\label{fig:Centauro_statistics}
\end{minipage}
\vspace*{3mm}
\end{figure}

The
experimental
 results  show that
 hadron-rich families constitute more than 20\% of the whole statistics
\cite{ref:Baradzei}
 as
 is
illustrated
in
Fig.~\ref{fig:Centauro_statistics}.
 This conclusion has been drawn from the analysis of  the unbiased sample
of
429 families
from Chacaltaya (open circles), 173 from the Pamir-Joint chambers
and 135 from a part of the Pamir chambers of 500 m$^{2}$yr (closed
circles) with  total visible energy greater than 100 TeV.    
 A scatter diagram of $N_{h}$ vs. $Q_{h}$ is shown, where $N_{h}$ denotes
the number
of hadrons in a family with visible energy greater than 4 TeV,
 and $Q_{h}=\Sigma E^{(\gamma)}_{h}/(\Sigma E_{(\gamma)}
 + \Sigma E^{\gamma}_{h})$ is the fraction of the total visible energy
 carried by these hadrons.

 The experimental data reveal the existence of several types Centauro
species such as:
\begin{enumerate}
\item{Centauros of original type,}
\item{Mini-Centauros,}
\item{Chirons,}
\item{Geminions.}
\end{enumerate}
 They are all  characterized by:
\begin{itemize}
\item Abnormal hadron dominance (both in multiplicity and in energy
   content). For ``classical'' Chacaltaya Centauros hadron multiplicity
$<N_{h}> \sim$ 75 in comparison to electron/gamma
      multiplicity 
$<N_{\gamma}>
      \sim$ 0.
\item Low total (hadron) multiplicity, in comparison with that
       expected for nucleus-nucleus collisions at that energy range.
\item {Transverse momentum of produced particles  much higher than
that
      observed in ``normal'' interactions ($p_{T} \approx$ 1.7 GeV/c
      for Centauros  and 10-15 GeV/c for Chirons,  assuming a gamma
      inelasticity coefficient $K_{\gamma}\approx$
      0.2).}
\item  Psedorapidity distributions  
        consistent with
a nearly
      Gaussian distribution and
 experimental characteristics supporting the formation and 
subsequent
      isotropic decay of a fireball with  hadron multiplicity
       $N_{h} \sim 100$ and  mass
 $M_{fb} \sim 180$ GeV for Centauros and Chirons, and
  $ N_{h} \sim 15$ and $M_{fb} \sim$ 35 GeV
       for Mini-Centauros. 
\end{itemize}

  Besides that they are
very frequently
 connected with the so-called {\em long-flying (penetrating) component}.
 In fact, the strongly penetrating
component has two aspects. At first, it has been observed in the apparatus
in the form of strongly  penetrating single  cascades, clusters of showers
or the
so called ``halo''. This phenomenon manifests itself  by the
characteristic
energy pattern revealed in shower development in the deep chambers
(calorimeters) indicating
the slow attenuation and many maxima structure.
The second aspect is connected with the anomalously strong penetrability
of some
objects in the passage through the atmosphere.
In principle, both aspects can be connected one to each other and could be   
 different manifestations  of the same phenomenon.

Anomalous cascade transition curves have been firstly noticed
during the study of Chiron-type families, where the simultaneous
appearance  
 of the unusual hadronic component with the short interaction
mean
      free path, as small as $\sim$ 1/3-1/2 of  the nucleon geometrical
      collision mean free path has been observed.
 Subsequently, the strongly penetrating component  has been encountered
also in
other kinds of events
\cite{ref:ewa}-\cite{ref:Baradzei}.
 Some of them penetrate through the
whole apparatus without noticeable attenuation, sometimes even indicating
 a tendency to grow.   
 The most  spectacular examples are two exotic cascades detected in the   
Centauro--like event C--K \cite{ref:buj} found in the homogenous type
 deep lead chamber.
 They were observed not far from the energy weighted
centre of  the family,
 at  very close distance one to each other. Both cascades
demonstrated a multicore
structure and unexpectedly  long
range and  many maxima character.
 After passing through a  very thick layer
of  lead, they escaped through the bottom of the chamber.
\begin{figure}[h]
\begin{center}
\mbox{
\epsfxsize=12cm
\epsfbox[-112 188 819
650]{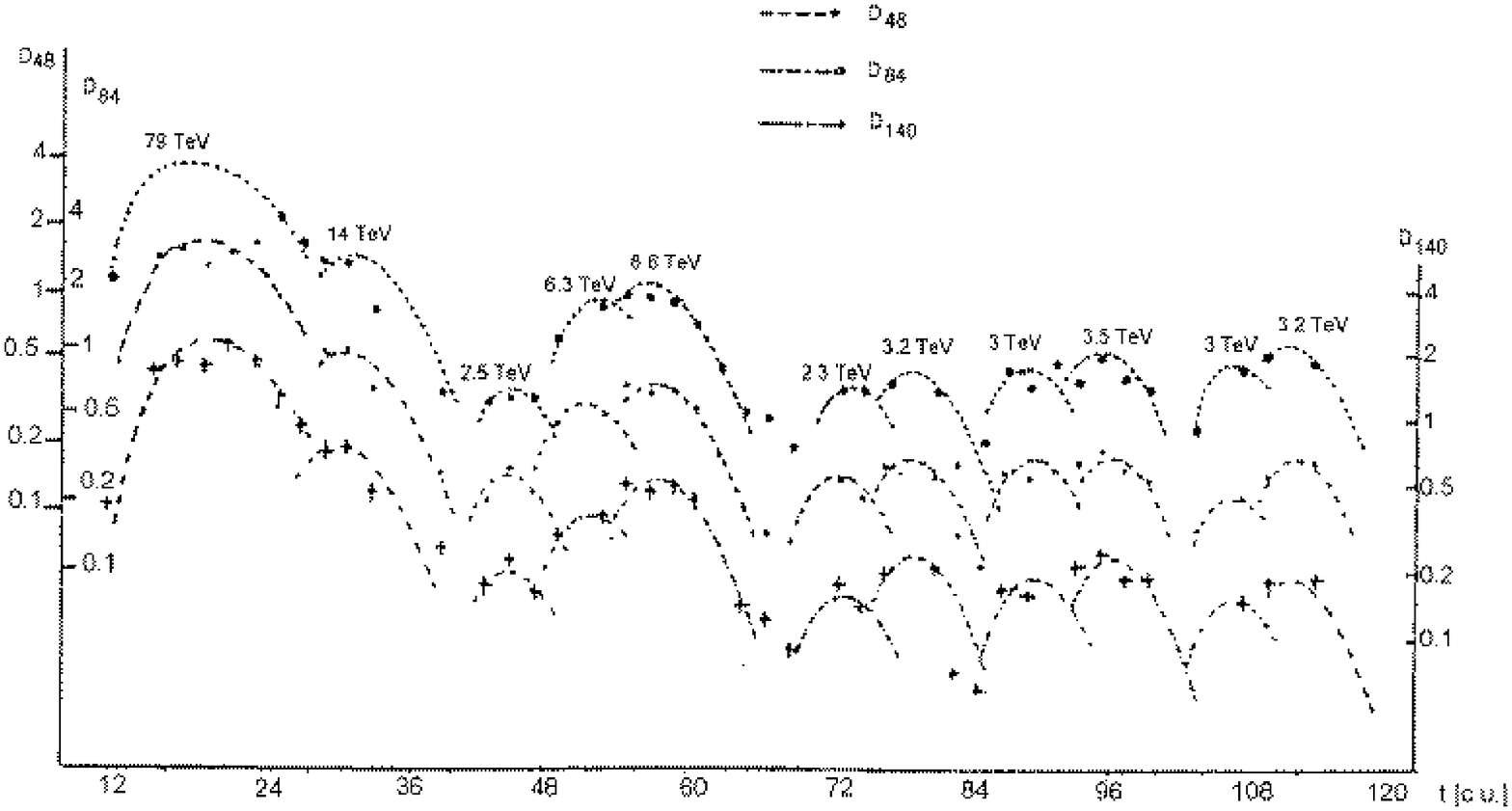}}
\caption[{\scriptsize Transition curves
 for cascade
no. 197.08 from Centauro C-K.}]
{\scriptsize Transition curves in X-ray film darkness D (measured
in
three diaphragms of\\ a radius R = 48, 84 and 140 $\mu$) for cascade
no. 197.08. Energy (in TeV units) liberated \\ into the
soft component is indicated at each hump (averaged over three estimated
values) \cite{ref:buj}.}
\label{fig:197}
\end{center}
\vspace*{-0.4cm}
\end{figure}
  The longer cascade, shown in
Fig.~\ref{fig:197}  penetrated more
than 109 cascade units and 11 maxima appeared along  its transition
curve.
The average distance between the neighbouring humps is about two times 
shorter than theoretically expected for conventional  hadronic cascades.
 Recent simulations,  assuming four  models
of hadron-nucleus interactions (VENUS 4.12, QGSJET, HDPM and modified UA5,
all widely used as standard models)  confirmed the unusual character
of long-penetrating cascades \cite{ref:tam}.

\section{Phenomenological model}
\label{sec:model}

\begin{figure}[h]
\begin{minipage}{7cm}
\begin{center}
\mbox{}
\epsfig{file = 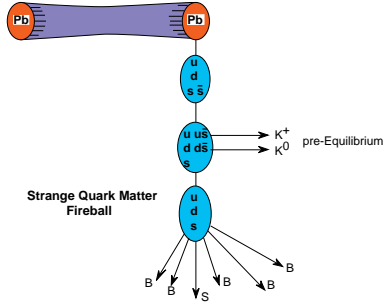,
bbllx=50,bblly=441,bburx=520,bbury=819,width=5cm}
\end{center}
\end{minipage}
\begin{minipage}{7.cm}
\vspace*{1cm}
\caption{{\scriptsize Centauro fireball evolution scheme.}}
\label{fig:Cen_evolution}
\end{minipage}
\end{figure}

According to the model developed in~\cite{ref:pa1,ref:asp}
Centauro arises through the hadronization of a QGP fireball of very high
baryochemical potential $\rm (\mu_b >> m_n)$, produced in nucleus-nucleus
collisions in the upper atmosphere.
In this model the QGP fireball initially consists of u and d quarks and
gluons. The very high baryochemical potential inhibits the creation of
$\rm u\bar{u}$ and $\rm d\bar{d}$ quark pairs, resulting in the fragmentation
of gluons predominantly into s$\bar{\rm s}$ pairs. In the eventual
hadronization of the fireball this leads to the strong suppression of pions
and hence of photons. As the fireball evolves strangeness distillation
through emission of kaons~\cite{ref:gre} transforms the initial quark matter
fireball into an increasingly strange quark matter state.
The stabilizing effects of the strangeness excess prolong the lifetime of
the fireball, enabling it to reach mountain-top altitudes~\cite{ref:the}.
In the subsequent decay and hadronization of this state separation of
strangeness can lead to the formation of non-strange baryons, hyperons
and/or
strangelets.
The main stages of the time development  of the Centauro fireball
 are shown in Fig.~\ref{fig:Cen_evolution}.

The hypothesis that
  strangelets can be identified as the
strongly penetrating particles frequently seen accompanying hadron-rich
cosmic ray events has been made and
checked~\cite{ref:asp,ref:gl1}.

\begin{figure}[h]
\vspace*{1mm}
\begin{minipage}{7cm}
\hbox{
\epsfxsize=180pt
\epsfbox[-211 151 690 695]{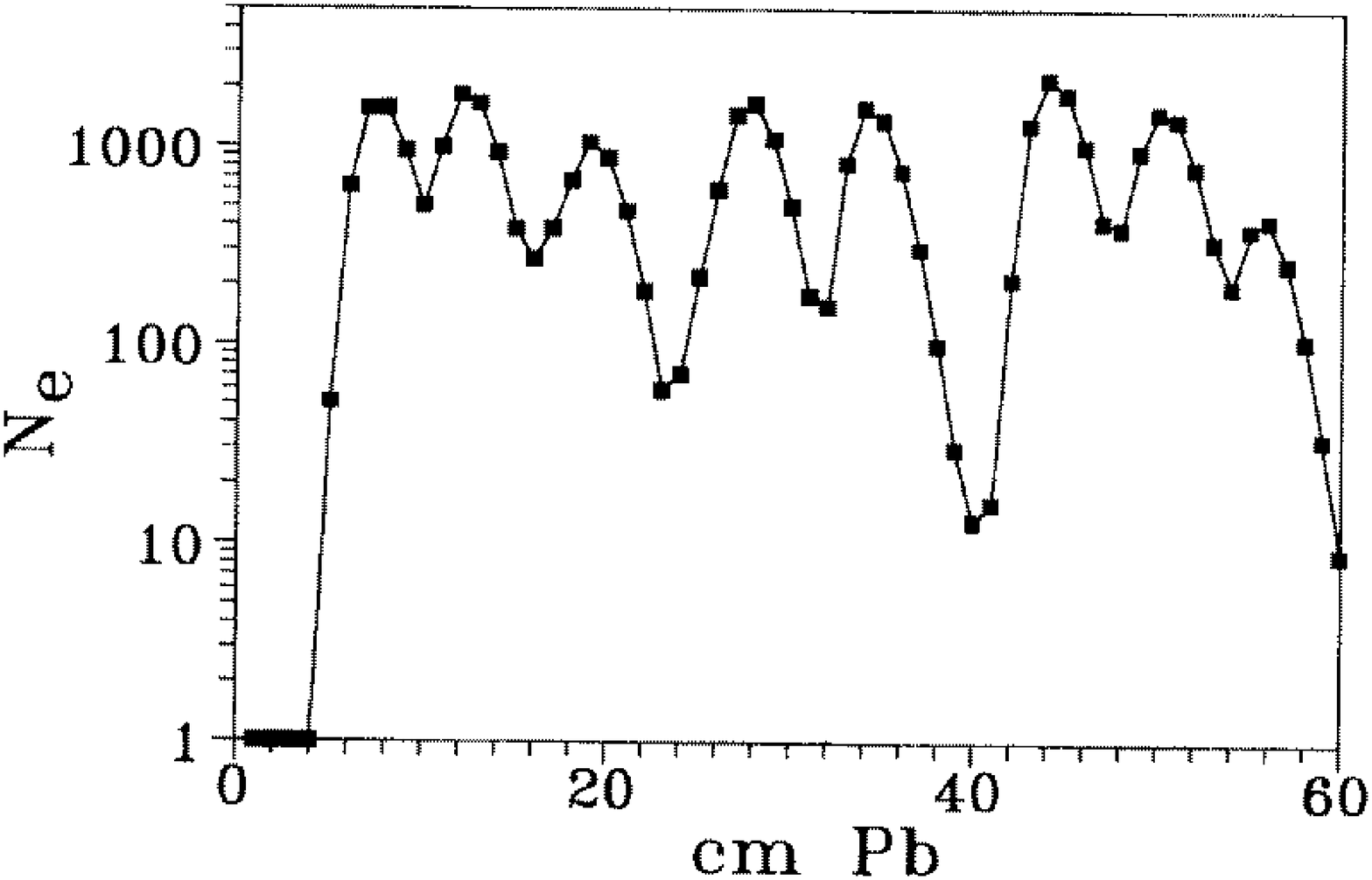}}
 \begin{center} {\scriptsize Unstable
 strangelet decaying into a bundle
 \vspace*{-1mm}\\ of 7 neutrons
 ($E_{n} \simeq  E_{str}/A_{str} \simeq
$
 200 TeV)}.
\end{center}
\end{minipage}
\begin{minipage}{8cm}
\vspace*{-4mm}
\hbox{
\epsfxsize=180pt
\epsfbox[-211 97 705 709]{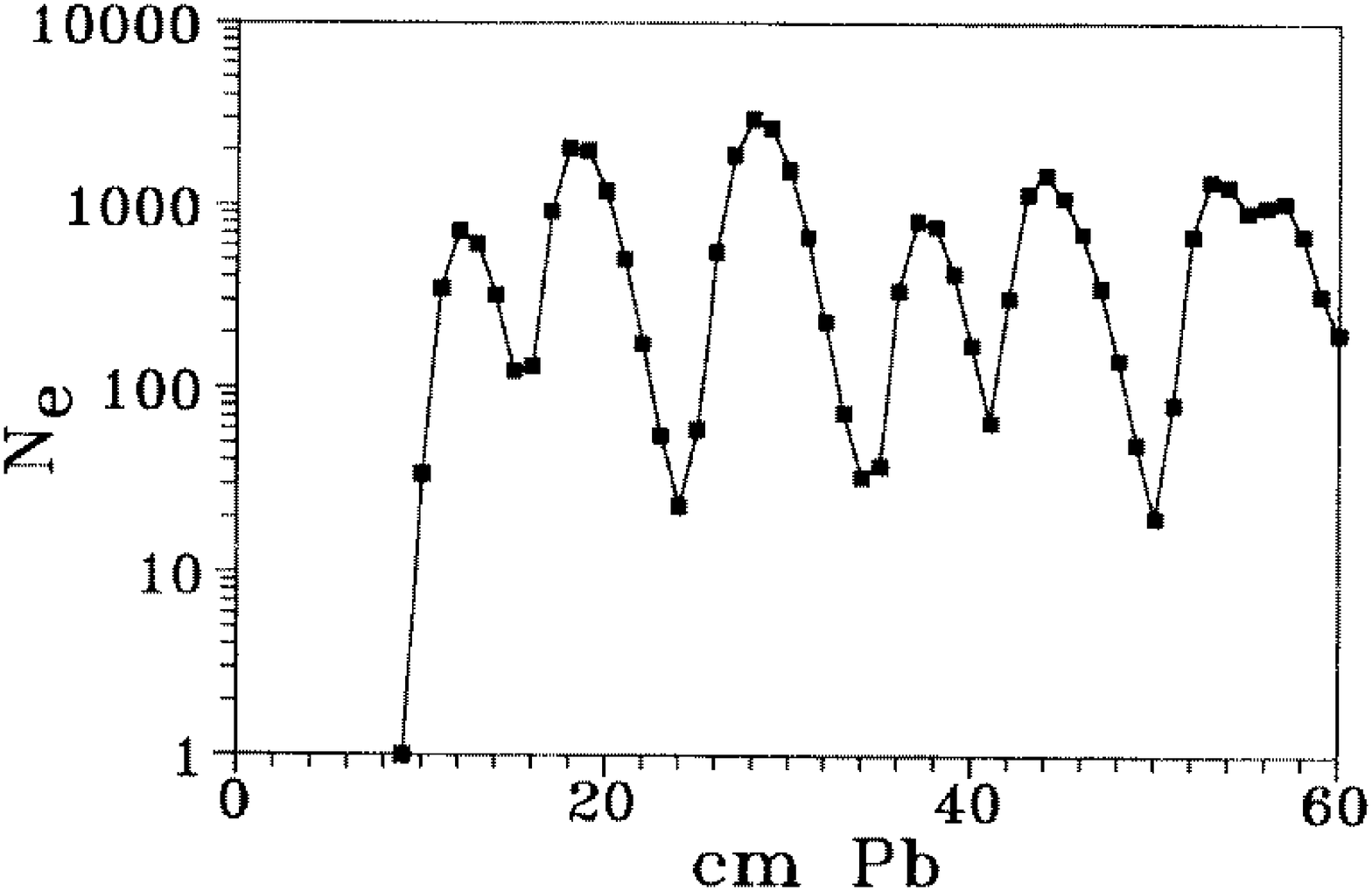}}
\vspace*{-1mm}
\begin{center} {\scriptsize Metastable strangelet\vspace*{-1mm}\\
  $(A_{str}=15, E_{str}=200$
 A TeV, $\tau \sim 10^{-15}$ s).}
\end{center}  
\end{minipage}
\begin{minipage}{7cm}
\vspace*{3mm}
\hbox{
\vspace*{4mm}
\epsfxsize=180pt
\epsfbox[-211 146 795
760]{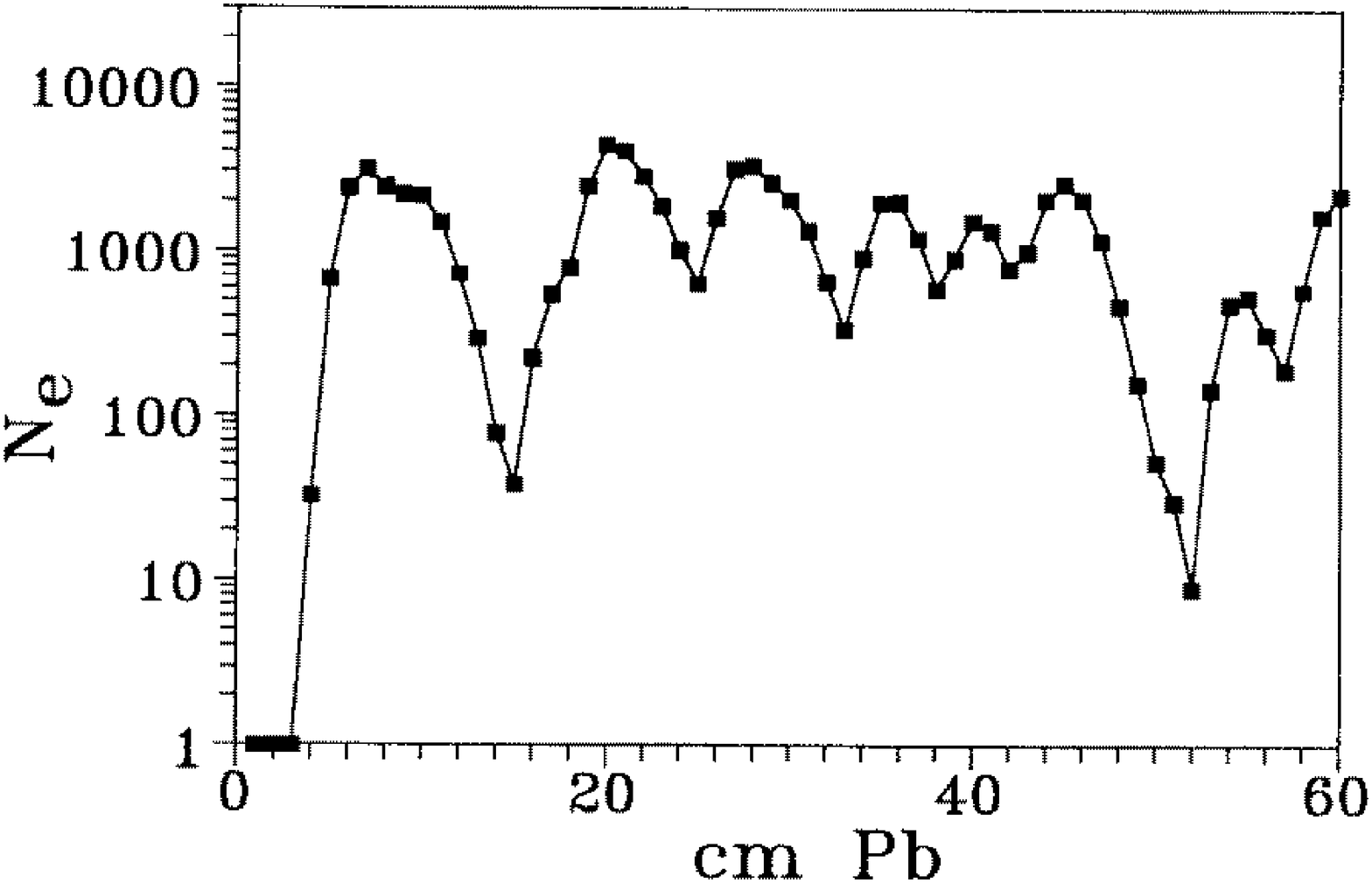}}
\begin{center}
\vspace*{-2mm}
\scriptsize{  
 Long-lived strangelet\\
 ( $ A_{str} = 15$, $\mu_{q} =$ 600 MeV).}
\end{center} 
\end{minipage}
%\hspace*{1cm}
\vspace*{-0.8cm}
\begin{minipage}{6cm}
\caption[{\scriptsize Examples of simulated transition curves recorded
in the lead chamber and produced by various strangelets.}]
{{\scriptsize Examples of simulated transition curves recorded
in the lead chamber and produced by various strangelets. Numbers of
electrons $N_{e}$ are counted within the
radius of
 50 $\mu$m.}}
\label{fig:stable_str}
\end{minipage}
\vspace{5mm}
\end{figure}

Our simulations show that the transition curves produced by strangelets
during their passage through the chamber resemble the experimentally
detected ones.
 As an example three cascade curves produced by
unstable,
metastable and unstable strangelets respectively are shown in
Fig.~\ref{fig:stable_str}.
 The long-range
cascades
observed in
the thick
homogenous lead/emulsion chambers could be the result of a
 strangelet
penetration  through the apparatus. Their strong penetrating power can be
connected both with the small interaction cross section (in comparison
with nucleus of comparable A) of the strangelet
and with the big concentration of its energy in a narrow region of phase
space. This energy could be liberated into conventional particle
production in many consecutive evaporation or interaction acts.   
 In this way numerous hadron-rich families accompanied by highly
penetrating cascades, clusters or halo  could be explained by     
assuming the
same mechanism of the formation of a strange quark matter fireball and its
successive decay into predominantly baryons and strangelet(s).

 Therefore, in this model both the basic characteristics of cosmic ray
Centauro
(small multiplicities and extreme imbalance of hadronic to electromagnetic
content) and the strongly penetrating component are naturally explained.
 The model is used to make predictions for Centauro and strangelet
formation in Pb+Pb collisions at the LHC. In 
table~\ref{tab:comp} we compare
characteristics of Centauro and strongly penetrating objects 
 (``strangelets''), either experimentally observed or calculated within
the
context of the model, for cosmic ray interactions and for Pb+Pb interactions
at the LHC.

\enlargethispage{60pt}
\begin{table}[h]
\begin{center}
\vspace*{-10pt}
\caption{Average characteristic quantities of modeled Centauro and \\
         Strangelets produced in Cosmic Rays and expected at the LHC.}
\label{tab:comp}
\begin{tabular}{ccc}
\hline
{\bf Centauro}         & {\bf Cosmic Rays}         & {\bf LHC}               \\
\hline
{Interaction                               } & { ``Fe + N''             } & { Pb + Pb              } \\
{$ \sqrt{s} $                              } & {$ \gtrsim $ 6.76 TeV    } & { 1148 TeV             } \\
{ Fireball mass                            } & {$ \gtrsim $ 180  GeV    } & {$ \sim $ 500 GeV      } \\
{ Projectile rapidity $ y_{proj} $         } & {$ \geq $ 11             } & {  8.67                } \\
{ Lorentz factor $ \gamma $                } & {$ \geq 10^4 $           } & {$ \simeq $ 300        } \\
{ Centauro pseudorapidity $ \eta_{cent} $  } & {     9.9                } & {$ \simeq $ 5.6        } \\
{$ \Delta\eta_{cent} $                     } & {     1                  } & {$ \simeq $  0.8       } \\
{$ <p_T> $                                 } & {    1.75 GeV            } & {   1.75 GeV (*)       } \\
{  Lifetime                                } & {$ 10^{-9} $ s           } & {$ 10^{-9} $ s (*)     } \\
{Decay probability                         } & {(x $ \geq $ 10 km) 10 \%} & {(x $ \leq $ 1 m) 1 \% } \\
{  Strangeness                             } & {  14                    } & { 60 - 80              } \\
{$ f_s $ (S/A)                             } & {$ \simeq $ 0.1-0.4 
} & {0.30 - 0.45           } \\
{   Z/A                                    } & {$ \simeq $ 0.3-0.4
} & {$ \simeq $ 0.2        } \\
{Event rate                                } & {$ \simeq $ 1 \%         } & {$ \simeq $ 0.1 \%     } \\ %  1000 events/ALICE-year} \\
\hline
{\bf ``Strangelet''                        } & {\bf Cosmic Rays}          & {\bf LHC}                \\
\hline
{  Mass                                    } & {$ \simeq $ 7 - 15 GeV   } & {10 - 80 GeV           } \\
{ $ f_s $                                  } & {$ \simeq $ 1            } & {$ \simeq $ 1          } \\
{ Strangelet pseudorapidity $ \eta_{str} $ } & {$ \eta_{cent} + 1.2 $   } & {$ \eta_{cent} + 1.2 $ } \\
\hline
\hspace*{-30mm} (*) assumed
\end{tabular}
\end{center}
\vspace*{-2mm}
\end{table}

\section{Strange Quark Matter - a possible state of matter}
\label{sec:str}

 As the search for strongly penetrating species which could be produced
by strangelets is one of the aims of the CASTOR detector, some questions
concerning their possible production and consequences of their existence
should be considered. It is especially important  in view of the
recent debate \cite{ref:Desai}-\cite{ref:Madsen2} on the possibility of a
catastrophic scenario
 initiated by
strangelets
formed in the future collider experiments.
 
The strange quark matter (SQM) is a matter with strangeness per baryon of
 the order
 of unity,  containing a  comparable fraction of up, down and strange 
quarks.  The existence of stable SQM was postulated by E. Witten
 \cite{ref:Witten}.
 Introduction of
a third flavour creates an additional Fermi well and thus reduces the
energy
relative to a two--flavour system what can thereby making the system
stable.
 The anticipated mass range for this kind of
matter lies anywhere between the masses of light nuclei and 
neutron
stars $A \simeq 10^{57}$. The latter  are called {\em strange stars}.
Strange matter as  part of cosmic radiation is  sometimes reffered to as
 strange quark {\em nuggets} or {\em nuclearities}.
 Smaller amounts of strange quark matter are
usually called {\em droplets} of strange quark matter, or simply
{\em strangelets}.
 Its existence, however,
is purely an experimental question since computations in QCD cannot come
close to answering this enormously important conjecture. SQM could have
important cosmological consequences for today's universe which arise 
from the possibility that the early universe might have undergone  a
first-order
phase transition from SQM to the nuclear matter.
 Strangelets left over from that era could also account 
for
the cosmological dark matter problem.

The
properties of some forms of hypothetical strange matter, as small
lumps of strange quark matter (strangelets) or hyperon matter (metastable
multihypernuclear objects MEMO's)  have been widely discussed  with special
emphasis
on their relevance to the present and future heavy ion experiments.
Different aspects of strange quark matter physics, as for example the
stability question  are described in the   
 reviews \cite{ref:stran_rev}.
Contrary to normal
nuclei, SQM stability increases with A and the threshold of its stability
is close to $A_{crit} \sim 300$. But also 
 quite  small strangelets might gain stability due to shell effects.  
However, due to the lack of theoretical constraints on bag model
parameters and difficulties in calculating color magnetic interactions
 and finite size effects, experiments are necessary to  answer the
 question of the stability of strangelets.

 Strangelets may arise from various scenarios; they could
be formed in high energy nuclear collisions
 or they might be of cosmological origin, as
remnants of the cosmic QCD phase transition. Collisions of strange
stars could also lead to the formation of strangelets which could
contribute to the cosmic ray flux. E. Witten
\cite{ref:Witten} suggested the possibility of production of
small lumps of SQM, in
today's universe by quark(neutron) stars, in the process of their
conversion to more
stable SQM stars.
 Several types of  models are applied to describe the
strangelet
production in heavy ion collisions.
%\cite{Heinz,Greiner,Baltz,Braun-Munzinger}.
They can be classified into two categories, namely {\it strangelet
production
by  coalescense} of hyperons or {\it by production following  a creation
of
quark-gluon
plasma}.
In a  very popular
coalescense model, an ensemble of quarks, which are products   
of nucleus-nucleus collisions, form a composite state which fuses
to form a strangelet \cite{ref:Baltz}.
 Formation of quark-gluon plasma
is not needed in this scenario, as hyperons coalesce during the late stage
of the collision forming a doorway state for strangelet production.
 Such a scenario  favours the  production of low mass strangelets
 ($A \geq 10$ is rather unlikely) in
the
midrapidity region.
Thermal models  assume  that chemical
and
thermal equilibrium are achieved prior to final particle production.
Coalescense and thermal models usually predict lower strangelet cross
section than models postulating  a QGP as an intermediate state
 in a strangelet formation. The
 {\it strangelet distillation mechanism}
provides a possibility for producing more stable large
strangelets since the QGP would lose energy by meson emission, possibly
resulting in a strangelet of approximately the same A as the QGP droplet.

The present status of strange quark matter searches, both in cosmic
rays and
in accelerator experiments is presented by R. Klingenberg in his
 recent review \cite{ref:Klingenberg}.
The experimental situation concerning the existence of SQM
is quite intriguing.

 Many
 accelerator experiments are looking for strange quark matter in heavy ion
collisions.
They are mostly based on
such discerning properties of strangelets as an unusual charge to mass
ratio ($Z/A \ll 1$). The strange counterparts of
ordinary nuclei are searched for in high-energy collisions at Brookhaven
%\cite{BNL_str}
 and at CERN. 
%\cite{CERN_str}
 To date no experiment has published results indicating a
clear positive signal for strangelets, although some candidates have been
 announced \cite{ref:CERN_str}.
 Accelerator  experiments are able to set  upper limits on the
existence of strangelets in the range of sensitivity of the experiments 
 but
they are not able to answer the question concerning their existence. 
There are several definite reasons for this. The main ones are  that
the
experiments are sensitive only to metastable strangelets with proper
lifetimes greater than $\sim 5\cdot 10^{-8}$ s and that they look for
strangelets
 in the midrapidity region.
Production limits obtained in these experiments are
strongly dependent upon the production model assumed for lumps of strange
quark matter.

Searches for SQM have been made also on terrestrial matter,
%\cite{str_terr}
cosmic rays and
astrophysical objects.
 The searches resulted in low limits
for strangelets in terrestrial matter but on the other hand
% it was postulated that the millisecond {\it pulsar SAX
%J1808.4-3658} is a
%good candidate for a strange star. Also
 {\it two seismic events} with
the properties of the passage of SQM nuggets  through the Earth have
been
 reported recently \cite{ref:seismic}.
Morever,
 a lot of
cosmic ray anomalies   could be understood by assuming the
presence of strangelets in cosmic ray spectrum.
%\cite{8,9},\cite{Shaulov}-\cite{Ichimura}.
 These, reviewed in
\cite{ref:ewa,ref:Wlodarczyk_rev} are:
\begin{enumerate}
\item
 {\it Massive and relatively low charged objects, i.e.}:
\begin{itemize}
\item Two anomalous events,  with value of charge $Z\simeq 14$
and  of mass number A$\simeq$ 350 and $\simeq$ 450
observed in primary cosmic rays by a balloon counter
experiment 
% (at the depth $\sim$ 9 g/cm$^{2}$)
 \cite{ref:Saito};
\item The so--called Price's event \cite{ref:Price_str}
% (detected at the depth $\sim$ 3-5 g/cm$^{2}$)
with Z $\simeq$ 46 and A $>$ 1000;
%regarded previously as a possible candidate for a magnetic monopole;
\item The so--called Exotic Track event with Z $\simeq$ 20 and
A $\simeq$ 460
      \cite{ref:Ichimura}. It was observed in a balloon-borne  emulsion
chamber
exposed
      to cosmic rays  at the atmospheric depth of only 11.7
g/cm$^{2}$ at zenith angle of 87.4 $^{0}$. This  means that the
projectile that 
caused that event traversed $\sim$ 200 g/cm$^{2}$ of the atmosphere.
 \end{itemize}
\item {\it Muon bundles from CosmoLEP \cite{ref:Cosmolep}}.
\item {\it Delayed neutrons}
\cite{ref:Zhdanov} observed
in  large Extensive Air Showers  by the neutron
monitor working in
conjunction with EAS instalation ``Hadron''.
\item {\it CENTAURO-related phenomena.}\\
 As well as Centauro-like events as the strongly penetrating component
are postulated to be the signs of strangelets passage through the matter.
Some explanations assume that strangelets are a component of primary
cosmic
rays \cite{ref:Shaulov},
 while others assume that they are produced in collisions of cosmic
ray
nuclei in the atmosphere \cite{ref:asp,ref:gl1}.
\end{enumerate}
  There are some  
  speculations \cite{ref:Desai,ref:Dar,ref:Busza} that
 under  some conditions the small metastable
SQM drops (produced for instance in  heavy ion collisions) could
be ''rapidly grown'' to a larger stable size, feeding in neutrons or 
light nuclei and releasing energy in photons. For such behaviour 
numerous conditions should be fulfilled. First of all strangelets  
 must  have rather low
velocity (must  be produced close to the midrapidity in collider
experiments) and a negative charge.  
 A number of
reasons have been put forward,
theoretical as well as experimental, why such a scenario is exceedingly
unlikely \cite{ref:Dar}--\cite{ref:Madsen2} and
 it is argued that  experiments at RHIC and LHC
do not represent a threat to our planet.
 The main arguments are:
\begin{itemize}
\item  Large safety factors
derived  from the survival of the Moon and the observed rate of
supernovae \cite{ref:Dar,ref:Busza};
\item Phenomena indicating the appearance of
 strangelets in 
cosmic rays;
\item Much higher probability of strangelet production in baryon--rich
      fragmentation regions than at midrapidity;
\item Predicted positive charge of strangelets
\cite{ref:Madsen1,ref:Madsen2}.
\end{itemize} 
 The strongest argument seems to be the existence of the
aformentioned 
phenomena which could be the manifestation of  strangelets
 in cosmic rays.
 Among theoretical arguments the prediction of a positive electric
 charge for strangelets seems to be especially convincing one.
 Strange matter 
 would have been neutral, if the ground state
composition consisted of equal numbers of quarks of three flavours,
 which is the most favoured state. Actually, however, the $s$ quark is
heavier than the other two flavours and so their number is slightly less
than that of the $u$ and $d$ quarks, as a result of which it has a
small positive charge.
J. Madsen  showed \cite{ref:Madsen1} that, in fact,
 stable SQM may be negatively charged in bulk, however, the
reduction in strange quark occupation in the surface layer,   
which is responsible for surface tension, causes the
intermediate mass strangelets to be always positively charged.
Morever, in his more recent work \cite{ref:Madsen2} he showed that
strange--quark matter in a color-flavour locked state is bound
stronger than 
``ordinary'' strange quark matter. This increases the probability  of 
 formation of metastable or even absolutely stable strangelets and leads
 to a stronger reduction in  number of negatively charged s-quarks in the
surface layer. A total positive quark charge of color-flavour locked
strangelets is expected to be proportional to the surface area or
 A$^{2/3}$. The
effect is large enough to rule out a potential disaster scenario.

\section{The CASTOR detector}
\label{sec:det}
\begin{figure}[h]
\vspace*{-2cm}
\begin{center}
\vspace*{-8pt}
\parbox[t]{0.48\hsize}{\epsfxsize=\hsize
%  \epsfbox{fig1.eps}
\epsfbox{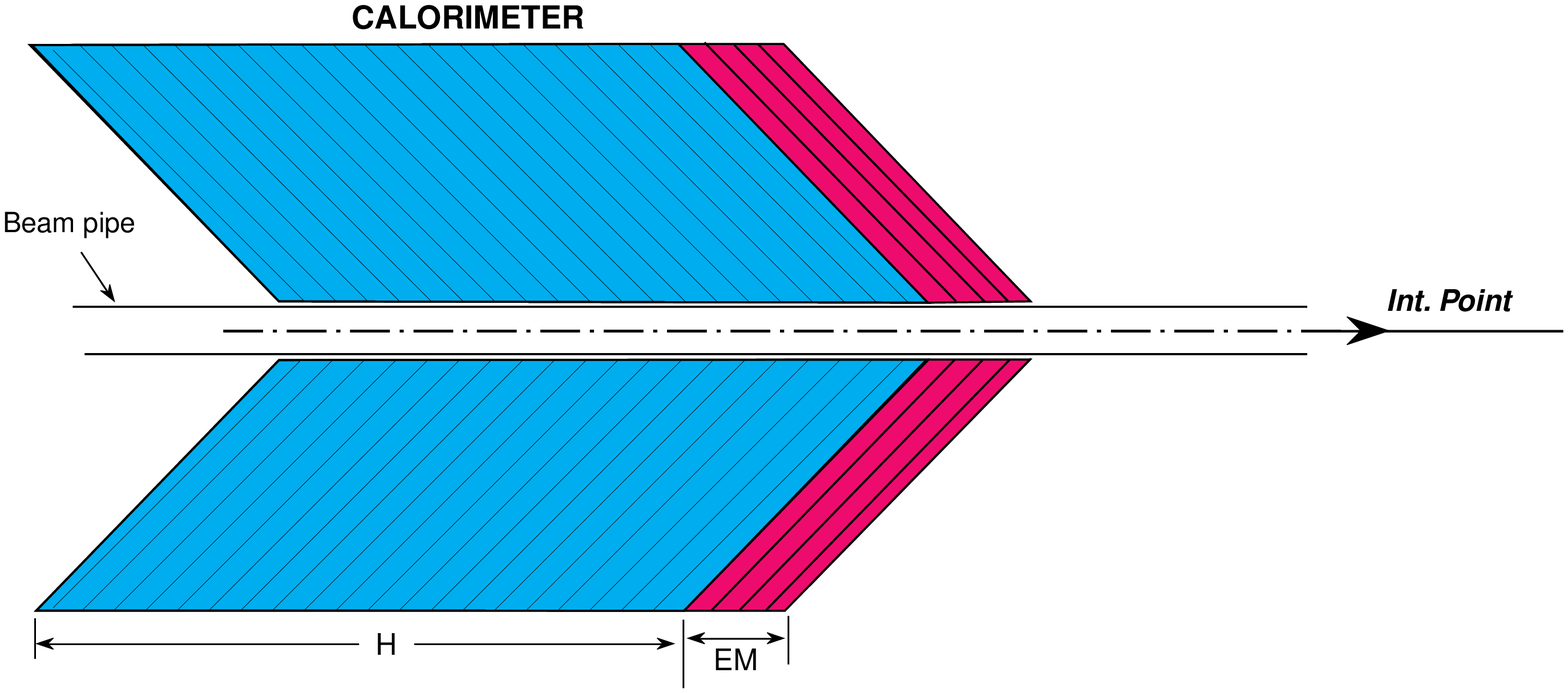}
{\vspace*{-22pt}}
 \vspace*{3mm}\caption[]{\vspace*{1mm}\\\hspace*{-2.5cm}Schematic side
view of 
the CASTOR
calorimeter.}}
    \label{fig:fig1}
\hfill
\parbox[t]{0.40\hsize}{\epsfxsize=\hsize
  \epsfbox[1 30 610 714]{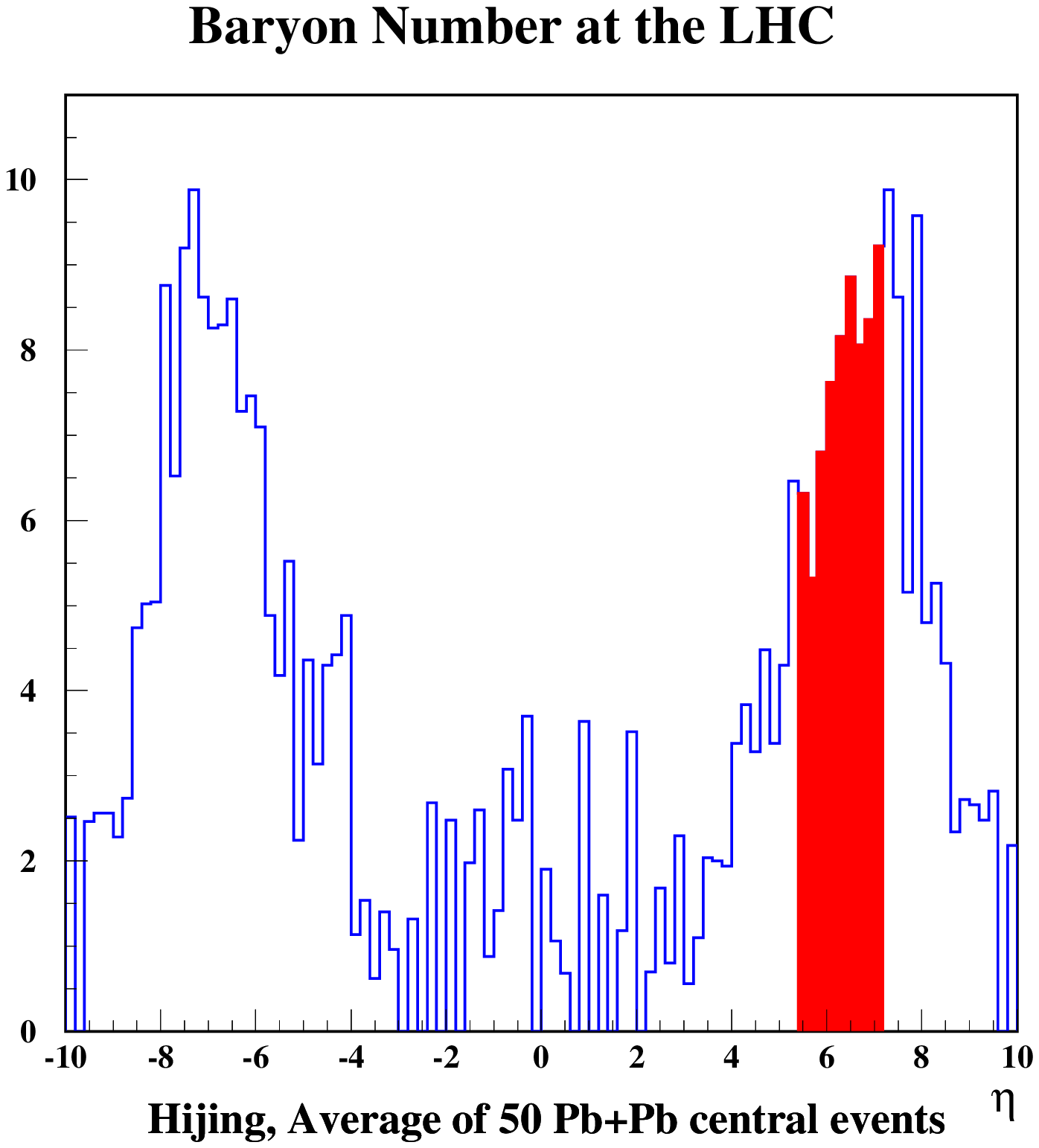}{\vspace*{-22pt}}
 \vspace*{2mm}   \caption[]{\vspace*{1mm}\\\hspace*{-3.0cm}
Baryon number pseudorapidity distribution pre-\\ \hspace*{-3.cm}
 dicted by HIJING for central Pb+Pb collisions\\\hspace*{-3.0cm}
 at the LHC, the CASTOR acceptance is indicated.
        }
    \label{fig:fig2}}
\end{center}
\end{figure}

CASTOR will probe the very forward rapidity region in nucleus-nucleus 
collisions
at the LHC and search for new effects. The main part of the detector will
be a deep calorimeter. Other parts, as for example photon and hadron
multiplicity detectors are also considered. The
calorimeter~\cite{ref:Castor} 
 (Fig.~\ref{fig:fig1})
 will be azimuthally divided into 8 octants and longitudinally segmented
into
layers, each layer consisting of a tugsten absorber plate followed by a
number
of quartz fibre planes. The signal is the Cherenkov light produced in
the fibres as they are traversed by relativistic charged particles
in the showers. The layers are inclined at 45$^o$ relative to the
impinging particles in order to maximize the Cherenkov light yield.
The light propagates along the fibres to the outer edge of the calorimeter
where it is collected by air light guides and transmitted to PMTs.
The detector will be 10 $\lambda_i$ deep and will be placed at 17 m
from the interaction point to cover the pseudorapidity range
$5.6 \lesssim\eta\lesssim 7.2$ where the baryon number density is expected
to be
large (Fig.~\ref{fig:fig2}).

%\enlargethispage{60pt}

 It will search for events with large
imbalance in electromagnetic and hadronic content and for abnormally
penetrating objects.

It was originally designed  and approved  as a subsystem of the ALICE 
heavy ion experiment. Because of technical reasons it has been recently
 moved to the CMS experiment.
Prototypes of the calorimeter have been constructed and tested with
 an electron  beam
at CERN. The last CASTOR calorimeter test took place in October 2001.
The preliminary results and comparison with the NA52 and H1 calorimeters,
employing
 similar technology are promising.

%\vspace*{-10pt}
\section{Simulations of exotic events}
\label{sec:sim}
\subsection{Centauro events}
In order to illustrate the detector's sensitivity to new effects we have
written the Monte-Carlo event generator CNGEN~\cite{ref:cng} embodying
the described model, and used it to study the production of Centauro in
the
forward, baryon-rich environment of a QGP fireball created in central
Pb+Pb
collisions at the LHC, and the detector's
response
to them.
 The results of simulations of Centauro events formed in Pb+Pb
central collisions at
$\sqrt{s}$ = 5.5 A TeV show 
that the
 characteristics  of simulated Centauro events are apparently different
 from
those  obtained from  conventional  (e.g. HIJING) generators
\cite{ref:ewa,ref:cng,ref:mix}.

\begin{figure}[h]
\hbox{
\epsfxsize=133pt
\epsfysize=140pt
\epsfbox[43 150 536 658]{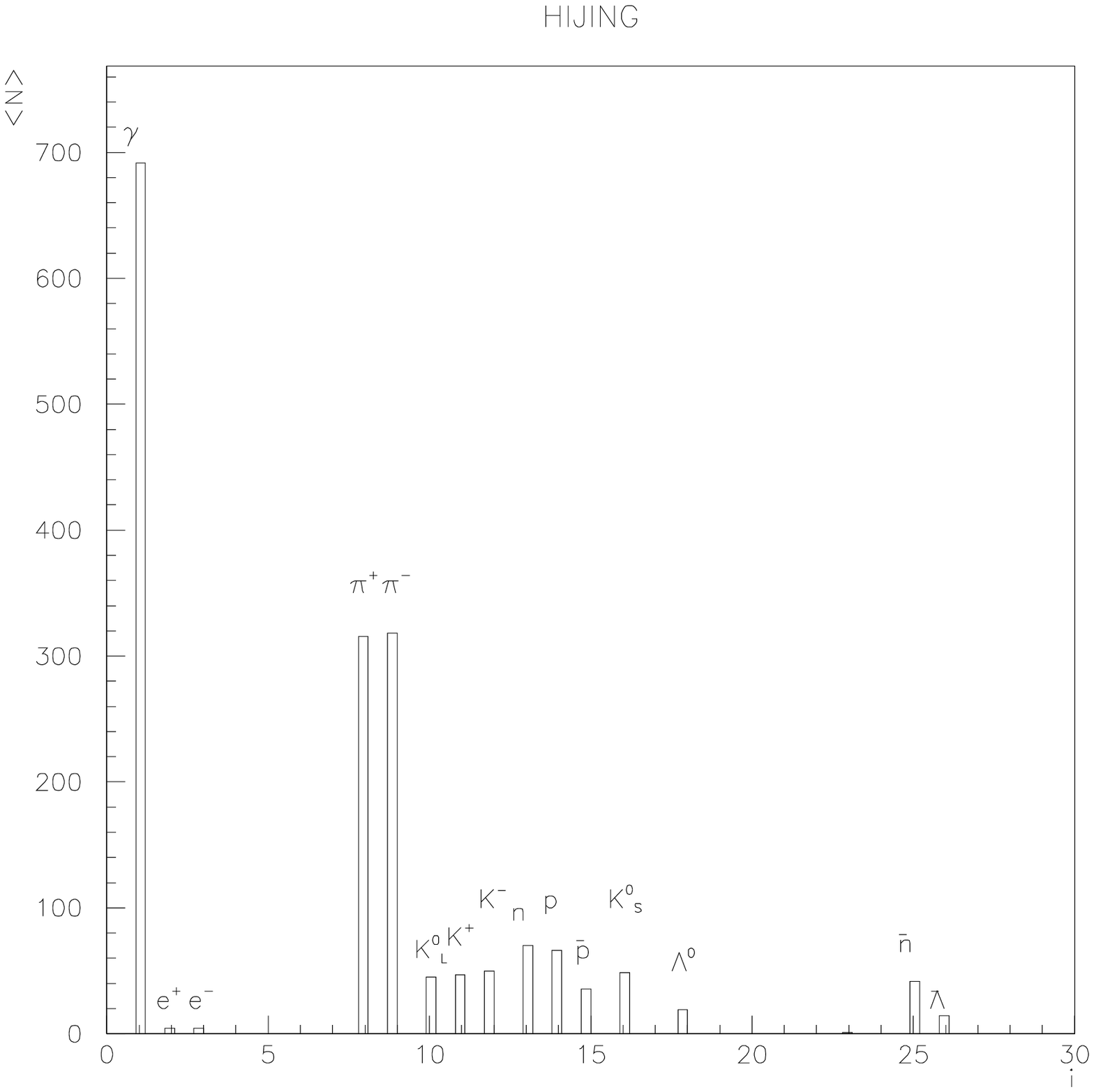}
\epsfxsize=133pt
\epsfysize=140pt
\epsfbox[43 150 536 658]{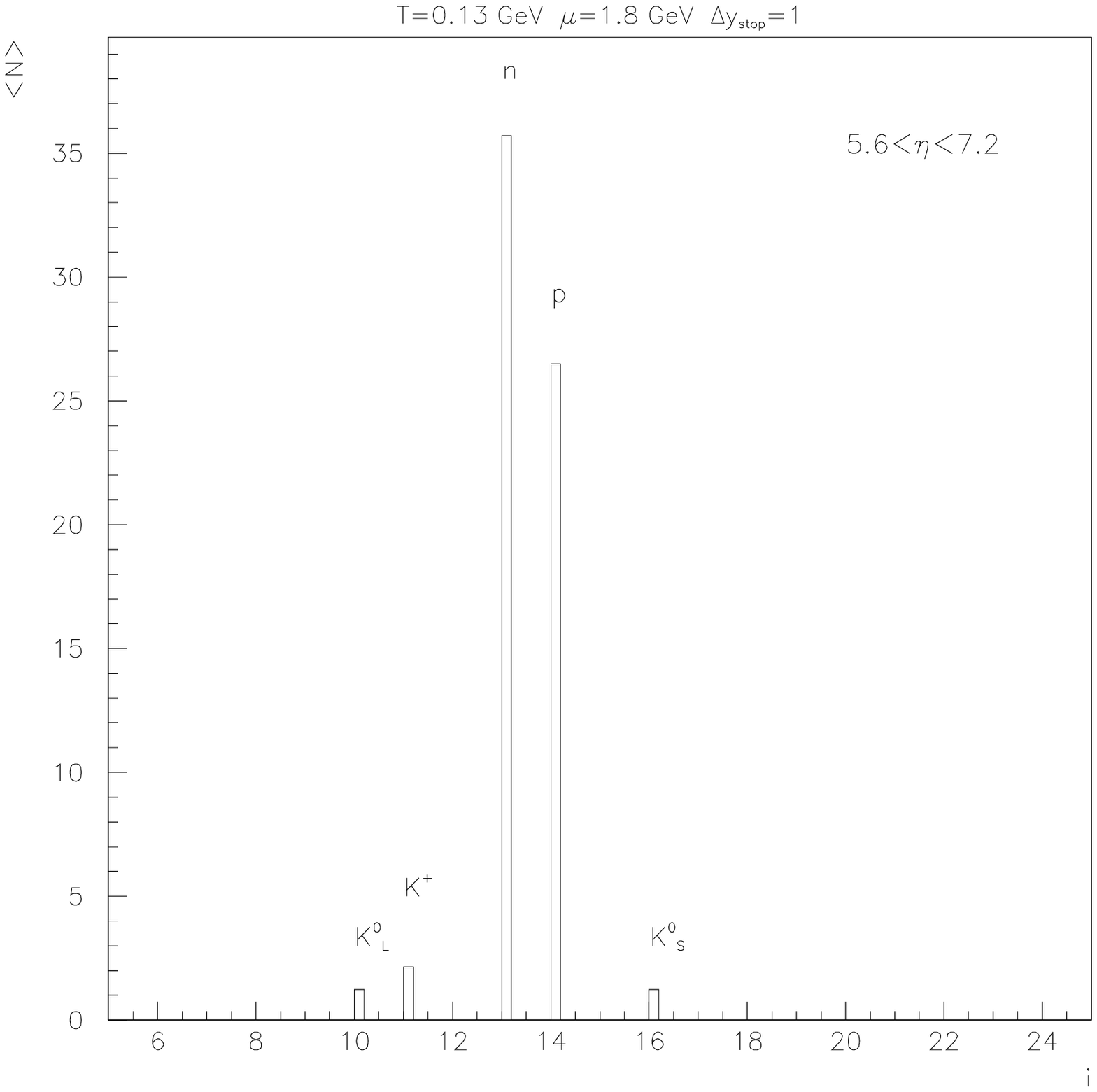}
\epsfxsize=133pt
\epsfysize=140pt 
\epsfbox[43 150 536 658]{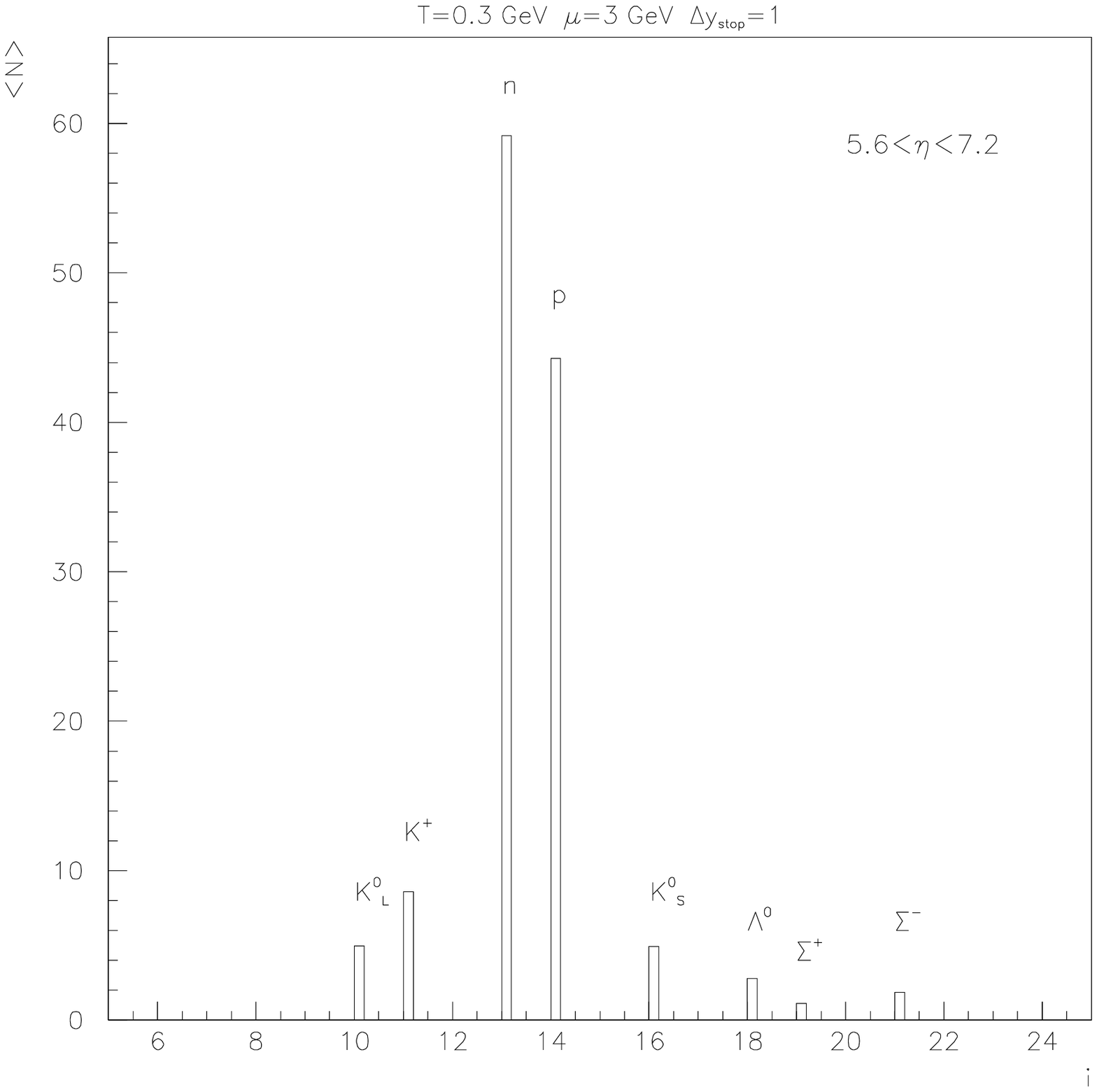}}
\vspace*{-3mm}
\caption[{\scriptsize Average multiplicities of particles produced in
 conventional 
 (HIJING) event and via Centauro mechanism.}]
 {\scriptsize{ Average
multiplicities of particles produced in
 conventional 
 (HIJING) event\\ and via Centauro mechanism. Only particles
 within CASTOR acceptance are shown.}}
\label{cen_hij}
\end{figure}

In particular, Centauro events are characterized by almost total absence
of the photonic component among secondary particles. The majority of
secondary particles are baryons. Kaons which are emitted from the primary
fireball can decay in principle  into neutral pions which in turn give
photons, but the overall  neutral pion production is suppressed here
strongly.
Small multiplicity of Centauro events may be surprising for
  nucleus-nucleus collisions at such energies.
It is illustrated in
Fig.~\ref{cen_hij} which  shows the average multiplicities of
different kinds
of particles produced via the  Centauro mechanism and contained within the
geometrical Castor acceptance.
 There are shown, as the examples,  the average
multiplicities for two sets of 
events
characterized by a temperature T = 130, 300  MeV and a
baryochemical potential $\mu_{b}$ = 1.8, 3.0 GeV
respectively, along with the  HIJING prediction for comparison.
 
\begin{figure}[h]
%\vspace*{-0.5cm}
\hbox{
\hspace*{2cm}
\epsfxsize=150pt
\epsfysize=150pt
\epsfbox[43 150 536 658]{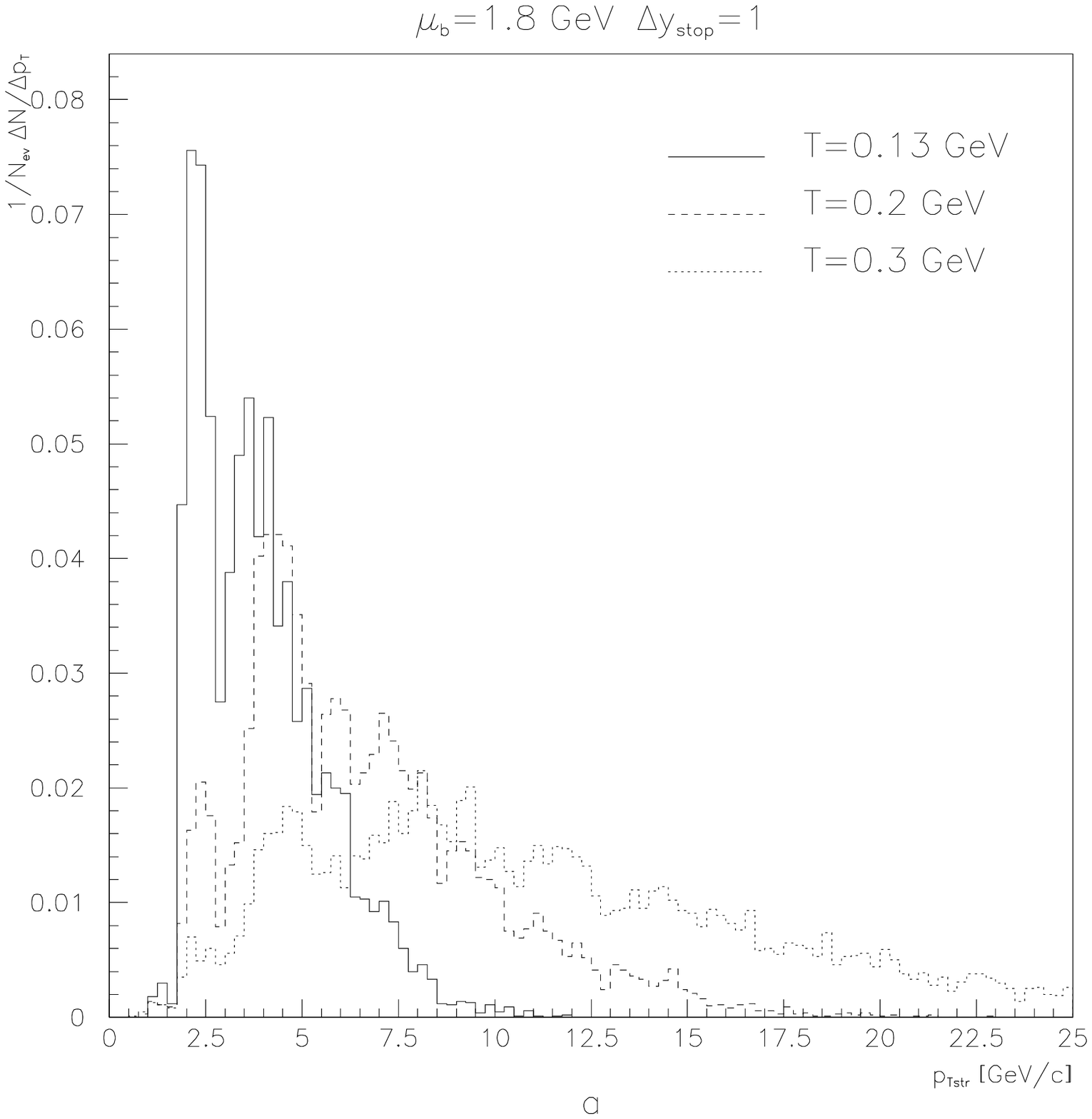}
\epsfxsize=150pt
\epsfysize=150pt
\epsfbox[43 150 536 658]{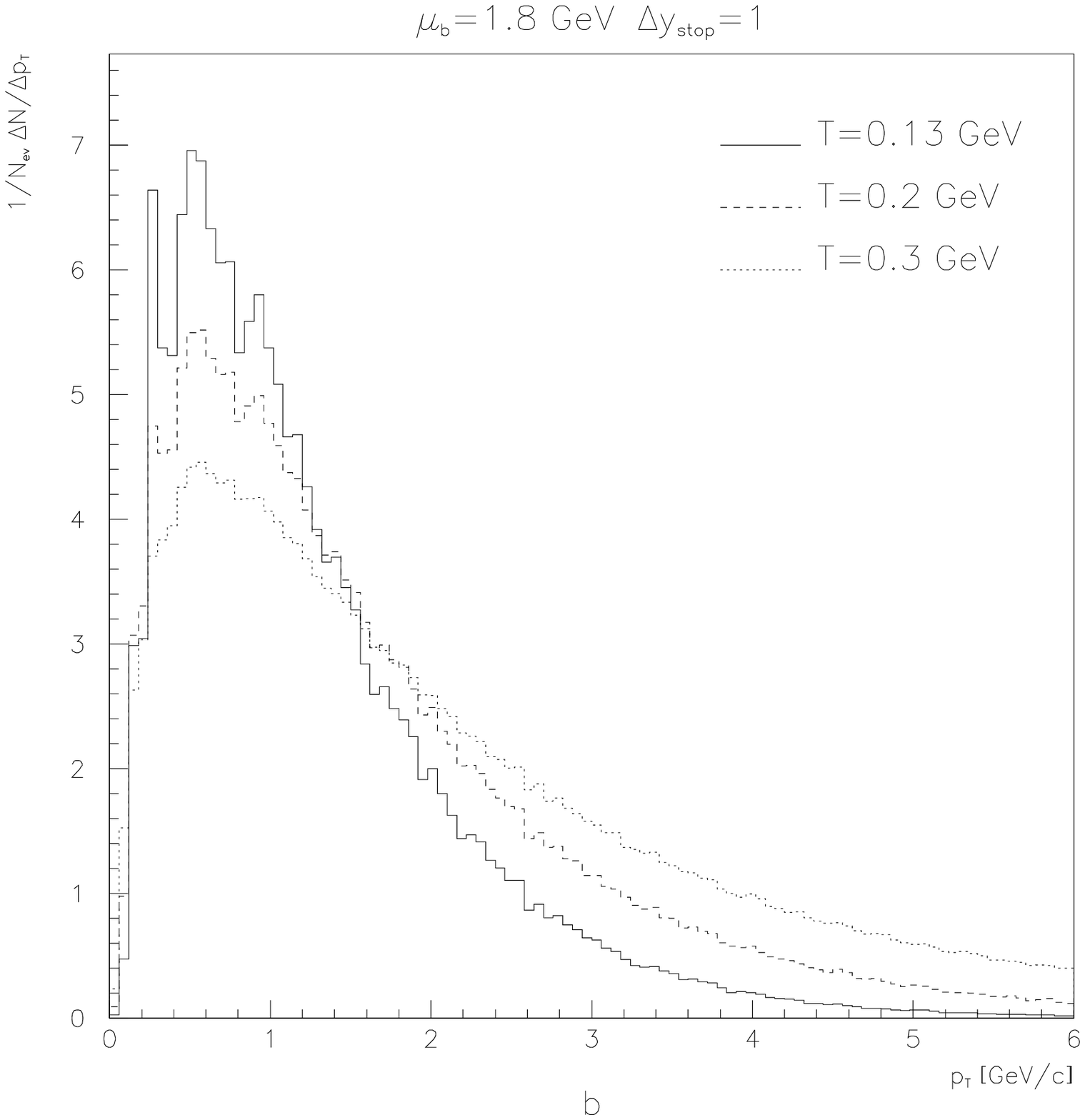}}
\caption{\scriptsize {Transverse momentum distributions of (a) strangelets
 and (b) hadrons from\\ Centauro decay.}}
\label{ptstr}
\end{figure}

Secondary particles in the Centauro events have larger mean  transverse
momentum in comparison with ordinary hadronic interactions. In
conventional 
events the average transverse momentum of produced particles $\langle
p_{T} \rangle$
= 0.44 GeV/c, as predicted by HIJING, which is several times smaller than
that of Centauro events. Fig.~\ref{ptstr} shows distributions of
transverse
momentum of strangelets and other Centauro decay products
 for three different temperatures T = 130, 200 and 300 MeV.

\begin{figure}[h]
\begin{center}
\vspace*{-8pt}
\parbox[t]{0.48\hsize}{\epsfxsize=\hsize
  \epsfbox{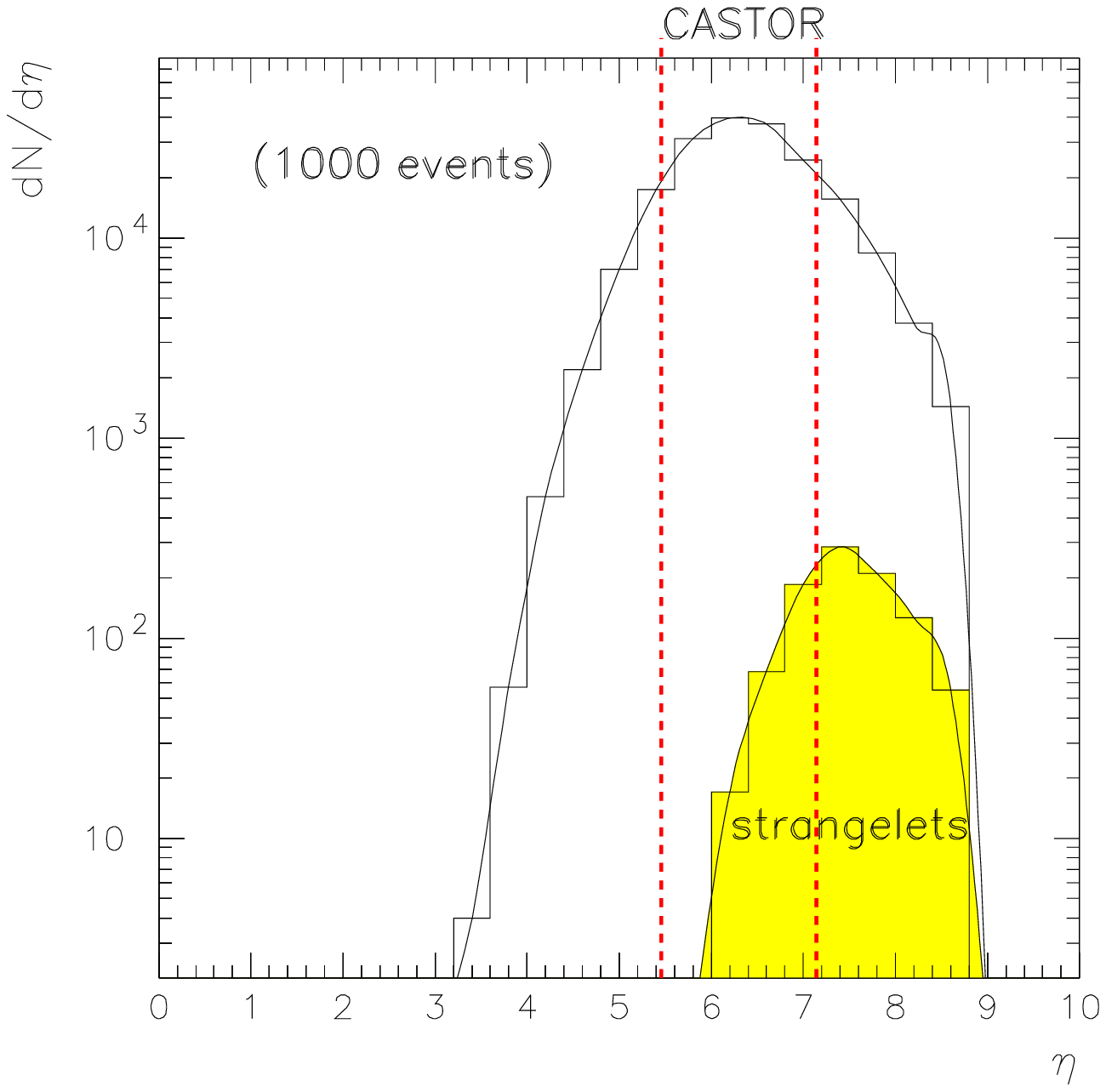}{\vspace*{-22pt}}
    \caption[]{\\\hspace*{-2.5cm}
                    Simulation of the production of Centauro and 
 \\ \hspace*{-2.5cm} strangelets in central Pb+Pb collisions at\\
   \hspace*{-2.5cm} the LHC, the CASTOR acceptance 
         is indicated.}
    \label{fig:eta}}
\hfill
\parbox[t]{0.40\hsize}{\epsfxsize=\hsize
  \epsfbox[1 30 610 714]{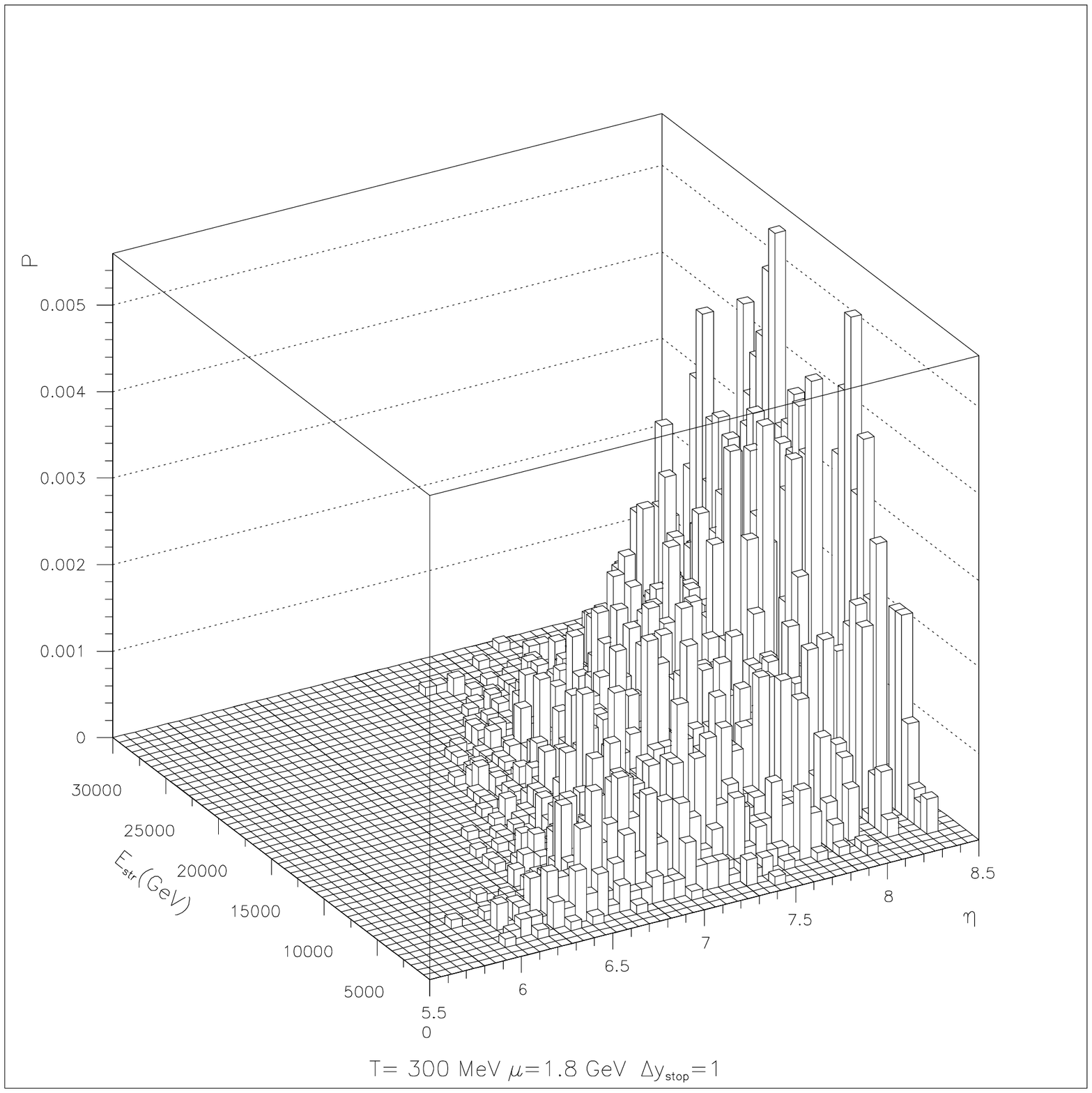}{\vspace*{-22pt}}
    \caption[]{\\\hspace*{-2.5cm}Probability of a strangelet production
as\\
        \hspace*{-2.5cm}a function of its energy 
        and pseudorapidity.}
    \label{fig:estr_eta}}
\end{center}
\vspace*{-0.5cm}
\end{figure}

The rapidity(pseudorapidity) distribution of decay products of the
Centauro fireball  depends mainly on the nuclear stopping power $\Delta
y_{stop}$.
Fig.~\ref{fig:eta} shows the distribution of particles from a decay of a
simulated Centauro fireball with T = 250 MeV and $\mu_b$ = 1.5 GeV.
A significant fraction of them falls within the CASTOR acceptance, as well
as $\simeq$ 25~\% of the associated strangelets.

%\enlargethispage{60pt}

Fig.~\ref{fig:estr_eta} shows two--dimensional lego histogram which
illustrates the probability of a strangelet production, as a function of
 pseudorapidity and energy.
 Most of these would have
E$_{str}$ = 5--10 TeV and A$_{str}$ = 20--40~\cite{ref:mix}
although
strangelets with much higher energy are also expected to be produced.

\subsection{Exotic objects in  deep  calorimeters}

The important question is what  signals will be produced by exotic
objects
during their passage through the deep calorimeter,
 and whether these signals can be distinguished
from those produced by conventional events.
 A possible background was estimated by means of the HIJING generator.
 Three  ``exotic'' scenarios were investigated. In the first one,
it
 was
assumed that  strangelets were born according to some undefined
mechanism anywhere  among other
conventionally produced
particles \cite{ref:aa2}.
 In the second one, strangelets  were considered to be 
   the
remnants of the Centauro
fireball explosion.
In this case the signal will be the sum of the strangelet signal
and the signal
 produced by nucleons coming from the isotropic
decay of the Centauro fireball
\cite{ref:mix}. 
%GEANT 3.21.
 In the third  scenario it was assumed that fluctuations of
electromagnetic
to hadronic
ratio were due to the production of a  DCC cluster.

\subsubsection{Strangelets\\}

At the LHC kinematical conditions,
 the
production of a variety of strangelets, characterized
by a wide spectrum of the  baryon number ($A_{str} \approx$ several tens,  
 temperature $T \approx$ 130-190 MeV and  quark chemical potential
$\mu_{q} \approx$ 600-1000 MeV) should be possible.  
 The
scenario, in which strangelets are born anywhere among other
conventionally
produced particles  was investigated in
\cite{ref:aa2} for both short-lived and long-lived strangelets.

  Unstable objects
which can decay via
strong interactions
 or the metastable ones decaying via weak
nucleonic decays were named {\bf short-lived strangelets}.
 If  lifetimes of metastable strangelets
are shorter than $\sim
10^{-10} $~s they could decay before reaching 
the CASTOR
calorimeter and give the  same picture  as  
the unstable
strangelets. The algorithm used in the calculations was the same
as the one used previously for cosmic ray events. It was assumed that
 unstable strangelets decay very fast, practically at the point of their
formation, thus the picture considered  resolves into the simple
case of a
bundle   
of neutrons entering the calorimeter.
The general conclusion concerning  the signals 
 produced in the
CASTOR calorimeter by
short-lived
 strangelets, formed in Pb+Pb interactions at the LHC
 is the same
as in the study  of the cosmic-ray strangelets. Bundles of
collimated  
neutrons can give in the CASTOR calorimeter an unconventional many-maxima
signal.

 The  objects
capable of reaching 
and
passing  through the CASTOR calorimeter  without decay, i.e.
 having a lifetime $\tau_{0} \geq  10^{-8}$~s have been named
{\bf long-lived
strangelets}.
Similarly, as in the case of cosmic ray strangelets,
 the simplified picture \cite{ref:gl1} of the interaction of a stable
strangelet
 in
the
calorimeter absorber was assumed.

 The strangelet was considered as an object with the radius
\[R = r_{0} A^{1/3}_{str}\]
where the rescaled radius is
\begin{equation}
r_{0} = ({\frac{3 \pi}{2(1 - \frac{2 \alpha_{c}}{\pi})[\mu^{3} + 
(\mu^{2} - m^{2})^{3/2}]}})^{1/3},
\end{equation}
$\mu$ and $m$ are the chemical potential and the mass of the strange quark 
respectively and $\alpha_{c}$ is the QCD coupling constant.
 The mean 
interaction path of strangelets in the tungsten absorber then is
\begin{equation}
\lambda_{s-W} = \frac{A_{W} \cdot m_{N}}{\pi(1.12 A^{1/3}_{W} + 
 r_{0}A^{1/3}_{str})^{2}}.
\end{equation}

 While penetrating through the calorimeter a strangelet collides with
tungsten
nuclei.
 At each  collision the spectator part of the strangelet survives
 continuing
its
passage through the calorimeter while  the wounded part is destroyed.
Particles
generated at the consecutive collision points  interact with the tungsten
nuclei
in the usual  way, resulting in the electromagnetic - nuclear
cascade
which
developes in the calorimeter.

\begin{figure}[hbt]
\begin{center}
\mbox{
\epsfxsize=280pt
\epsfbox[1 0 511 566]{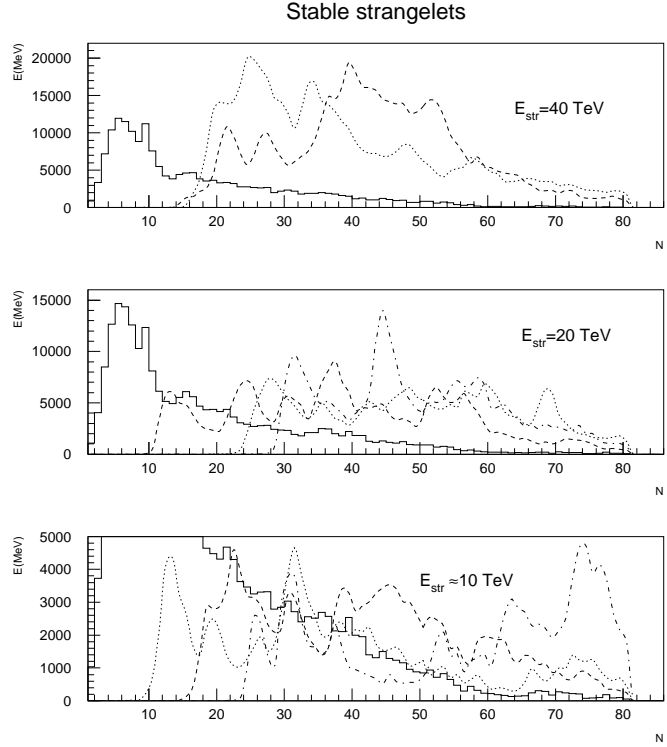}}
\end{center}
\vspace*{-1cm}
\caption [\scriptsize{
 Transition curves of stable strangelets.}]
{\scriptsize{
Transition curves of stable strangelets with energy $E_{str}$ = 10-40 TeV,\\
baryon number $A_{str}$ = 15-40, quark chemical potential $\mu_{q}$ = 600,
1000 MeV. Energy\\ deposit (MeV) in the calorimeter layers, in the
octant
containing a strangelet, is shown.\\ Full line histograms show the
HIJING
estimated background.}}
\label{stable_3}
\vspace*{-2mm}
\end{figure}

 Penetration of stable strangelets through the
calorimeter was simulated, assuming $\alpha_{s}$ = 0.3 and several
different sets of the
initial strangelet parameters:
 ($\mu_{q}$ = 300, 600, 1000 MeV;  
 $A_{str}$ = 15, 20, 40;
 $E_{str} \approx$ 8 - 40 TeV (or 400 - 1000
 $\times$ A GeV )).
Examples of transition curves produced in the CASTOR calorimeter by
various
stable strangelets are presented in
Fig.~\ref{stable_3}.
The
curves
are limited to the one  calorimeter octant which contains the  strangelet.
The
strangelet cascade profiles are compared with those of the conventional
background,
produced by particles generated by HIJING (full line histogram), after the
subtraction of the energy carried by the strangelet.
 It can been concluded that
 stable strangelets can produce in the calorimeter long range
many-maxima cascades,
 manifestly different from these produced by
 a conventional event.

\subsubsection{Strangelets and Centauros\\}

The other question concerns the shape of  transition
curves produced in the
calorimeter  by:
\begin{itemize}
\item   strangelets born in the Centauro fireball explosion
and  registered in the apparatus  together with other
Centauro
decay products,
\item or  Centauro fireball decay products  without accompanying
strangelets emission. 
\end{itemize}

\begin{figure}
\hspace*{1cm}   
\begin{minipage}{7cm}
\hbox{
\epsfxsize=180pt
\epsfbox[43 150 536
658]{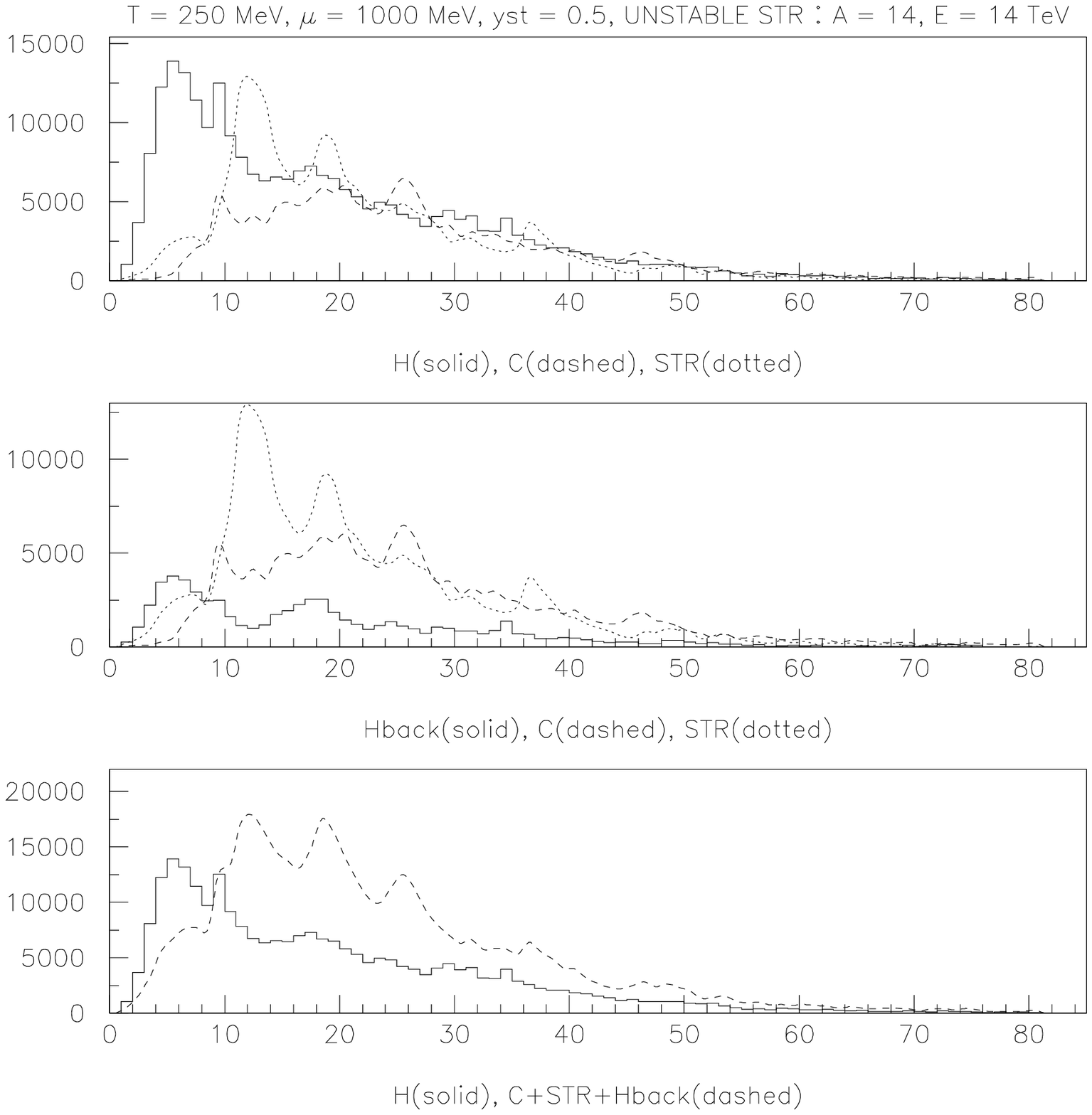}}
\end{minipage}
\begin{minipage}{7cm}
\hbox{
\epsfxsize=180pt
\epsfbox[43 150 536
658]{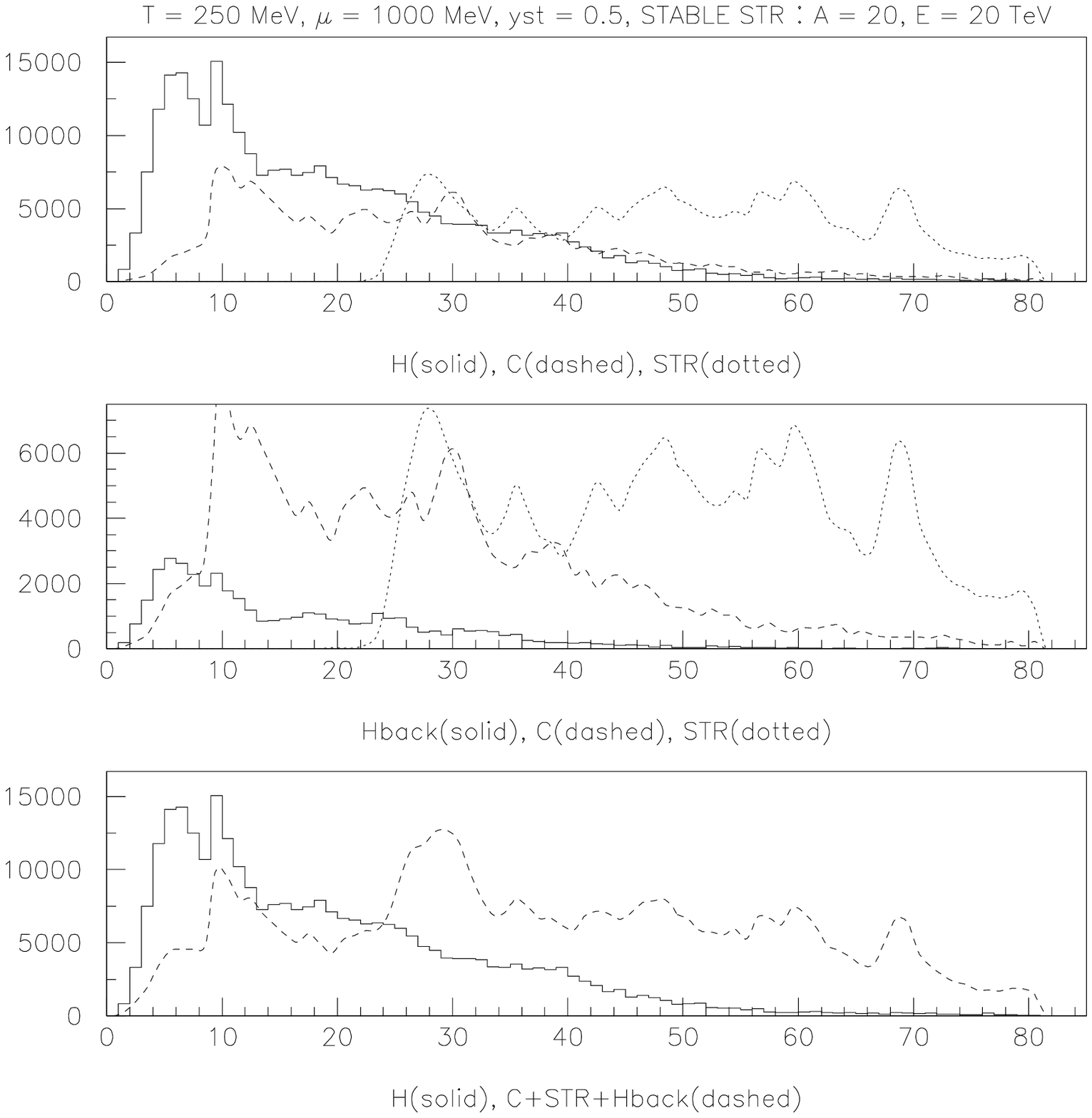}}
\end{minipage}
\caption[\scriptsize{
 Transition curves produced by Centauro event
   with unstable or stable strangelets, in comparison with
HIJING.}]
{\scriptsize{
Transition curves produced by Centauro event ``C''with unstable or\\
stable strangelets ``STR'', in comparison with HIJING ``H''. Energy 
deposit (MeV)\\ in consecutive  calorimeter layers, in the octant
containing
a strangelet, is shown.}}
\label{st_20_20_bis}
\vspace*{-2mm} 
\end{figure}

 Exotic events generated  
   by means of the Centauro code 
 were  passed  through the CASTOR
calorimeter,
by using a modified version of GEANT 3.21 \cite{ref:ewa,ref:mix}. For each
event, 
the  transition curves produced by
the Centauro fireball decay products,
 by the accompanying strangelet,
 and by background of conventionally produced
particles,
were simulated separately.
All these three contributions separately and also their sum,
 constituting the so--called ``mixed'' event,
were compared with the conventional transition curve, produced by HIJING.
Centauro events characterized by  various
values of 
parameters: temperature ($T$ = 250, 300 MeV), quarkchemical 
  potential ($\mu_{q}$ =
 600, 1000 MeV) and nuclear stopping power ($\Delta y_{stop}$ =
 0.5, 1.0, 1.5 corresponding  to an effective stopping in the range
  $\sim$ 1.5 - 3.0 pseudorapidity units) were analysed.
 In these events  strangelets
with baryonic numbers $A_{str}$ = 20-40 and energies
$E_{str} \simeq $ 8-20 TeV were formed.
 Two examples of the resulting transition curves produced by
 a Centauro event with an unstable and stable strangelet born among
its secondaries  are shown in
Fig.~\ref{st_20_20_bis}. The energies of exotic objects (Centauro and
 the strangelet) within the CASTOR acceptance were: $\sim$ 128 TeV and
$\sim$ 126 TeV respectively in comparison to $\sim$ 156 TeV predicted 
 by  HIJING for the conventional central Pb+Pb event.

 Our analysis indicates that the Centauro events (whether  accompanied
and not accompanied by a strangelet) can be easily distinguished from  
 conventional events. Centauro transition curve in the calorimeter
is expected to have manifestly  different shape and longer extent
than that produced by a  conventional event.
  The Centauro produced signal
has a maximum at about 14th calorimeter layer with
the
average at $\langle N_{Cent} \rangle \simeq$ 25. The HIJING event produces
the
maximum of the signal at about the  8th calorimeter layer, with the
average
at $\langle N_{HIJ} \rangle \simeq $ 19.
 Generally, the Centauro produced
signal is stronger in the deeper (hadronic)  part of the calorimeter,
  as opposed to the HIJING generated one, which is peaked in the
electromagnetic section of the calorimeter.

\subsubsection{DCC\\}

The QCD phase transition from normal hadronic matter to the Quark-Gluon-Plasma,
manifests itself in two forms: deconfinement transition and chiral
symmetry restoration.
 One of interesting consequences of the chiral
transition is the possible formation of a chiral condensate in an extended
domain,
such that the direction of the condensate is misaligned from the true
vacuum direction. Formation of these so called Disoriented Chiral
Condensate (DCC) domains in high energy collisions of both hadrons and
heavy
ions, has been proposed by many authors (see for
example \cite{ref:DCC_Bjorken}).
 A basic signature of DCC production is the presence of very
large
event-by-event fluctuations in the fraction of produced neutral pions.
Thus, there are some suspicions that this phenomenon could also be
responsible
for  the Centauro-like events. We have simulated the passage of DCC
clusters (both charged
and neutral) through the deep calorimeter and they also show 
 characteristic transition curves~(Fig.~\ref{fig:ex}).

\subsubsection{Summary of exotic events\\}

\begin{figure}[h]
\vspace*{-6mm}
\begin{center}
\vbox{
\epsfxsize=300pt
\epsfbox[1 121 584 703]{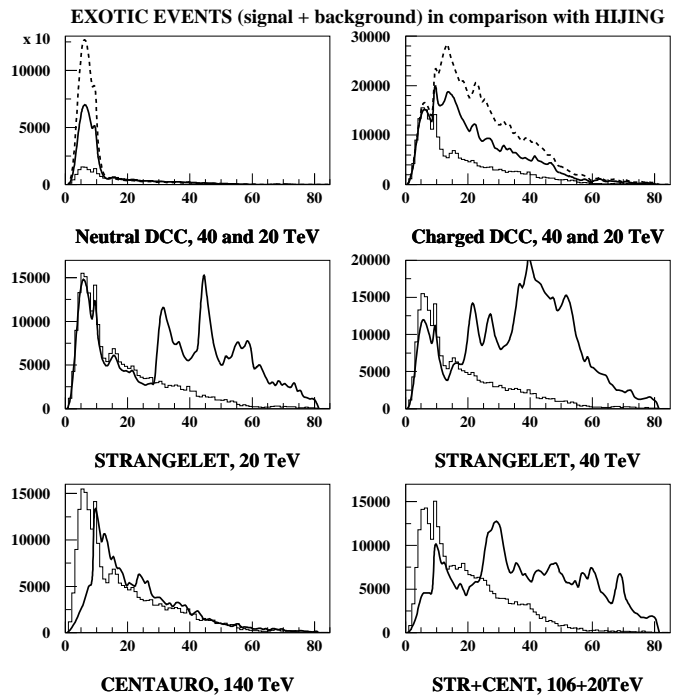}}
\vspace*{-3mm}
\caption{{\scriptsize Simulated signals produced by different kinds of
events. 
 The deposit of energy\\ in one sector (hit by the exotic object)
 vs.  layer number of the calorimeter is shown.\\ Thin continuous line
corresponds to the HIJING event.}}
\label{fig:ex}
\end{center}
\label{ex}
\vspace*{-5mm}
\end{figure}
 
In Fig.~\ref{fig:ex}
  are shown for comparison  examples of transition curves
produced
by different exotic phenomena, among them charged and neutral DCC
clusters,
strangelets, Centauros and mixed events. The  curves presented are the sum
of the exotic signals and the conventional background.
 All these phenomena give different
 patterns and could be easily distinguished  from  
 conventional  events as well as from each other. This method is
sensitive to the 
detection of any kind of strangelets: both neutral and charged, as well
 as short-lived and long-lived. This is an important feature  in the
 light of  present experiments which are only able to detect strangelets
with lifetimes longer than $\sim 10^{-9}$ s. It
is seen that the energy
deposition pattern in a deep calorimeter can be an excellent signature of
a QGP state decaying in different exotic ways.

\begin{figure}[h]
\vspace*{-4mm}
\begin{minipage}{6cm}
\epsfxsize=150pt
\epsffile{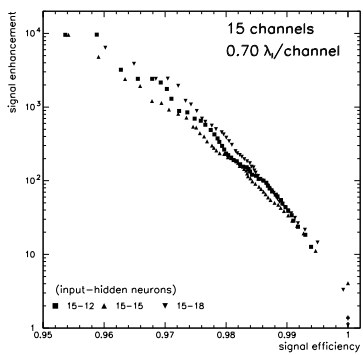}
\end{minipage}
\begin{minipage}{9cm}
\vspace*{1.cm}
    \caption[]{Enhancement of 
      the signal/background ratio for strangelet 
         detection obtained through application of 
         the neural network technique.}
    \label{fig:fig5}
\end{minipage}
\end{figure}

For the configuration of the calorimeter studied here the extraction of
the exotic phenomena signal  from the conventionally produced
background
will be straight forward  for energies higher that $\sim$ 10 TeV.   
In order to study the sensitivity of the calorimeter to abnormally
penetrating objects of lower energies, we developed a neural network
technique~\cite{ref:SQM2001}
based on a multilayer perception network. On the input layer are fed the
signals from the readout units of one octant, the output layer consists of
one neuron providing a yes/no answer, and there is one hidden layer.
We tested the method using simulated strangelets.
One set of 10000 5 TeV strangelets and 10000 background events was used
as the training sample and a different such set as the evaluation sample.
Figure~\ref{fig:fig5} shows the enhancement obtained in the signal/background
ratio through the application of this technique to a calorimeter
consisting
of 15 readout units per octant, for different numbers of neurons in the
hidden layer.
An enhancement of {\it O}(10$^4$) is obtained for efficiency $\simeq$
95~\%.

\vspace*{0.2cm}
This work was partly supported by Polish State Committee for Scientific 
Research grant No. 2P03B 011 18 and SPUB-M/CERN/P-03/DZ 327/2000.

\enlargethispage{60pt}

\end{document}